\documentclass[10pt]{article}

\usepackage{amsmath}
\usepackage{amssymb}
\usepackage{textcomp}
\usepackage{ulem}
\usepackage{upgreek}

\usepackage{graphicx}
\usepackage{nameref}
\usepackage{cite}

\usepackage[usenames,dvipsnames]{color}

\usepackage[labelfont=bf,labelsep=period,justification=raggedright]{caption}

\makeatletter
\renewcommand{\@biblabel}[1]{\quad#1.}
\makeatother

\date{}

\usepackage{epstopdf}

\begin{document}

\begin{flushleft}
\begin{center}
{\Large
\textbf{Correlations and functional connections\\
\vspace{0.2cm} in a population of grid cells}
\vspace{0.2cm}}\\
\end{center}
Benjamin Dunn$^{1,\ast}$, Maria M\o{}rreaunet$^{1,\ast}$, Yasser Roudi$^{1,2,\ast\ast}$ 
\\
$^1$ Kavli Institute for Systems Neuroscience and Centre for Neural Computation, NTNU, 7030 Trondheim, Norway \\
$^2$ Nordita, KTH and Stockholm University, 16903 Stockholm, Sweden \\
$\ast$ authors contributed equally \\
$\ast\ast$ E-mail: yasser.roudi@ntnu.no
\end{flushleft}
\section*{Abstract}
We study the statistics of spike trains of simultaneously recorded grid cells in freely behaving rats. 
We evaluate pairwise correlations between these cells and,  
using a generalized linear model (kinetic Ising model), study their functional connectivity. 
Even when we account for the covariations in firing rates due to overlapping fields, both the 
pairwise correlations and functional connections decay as a function of 
the shortest distance between the vertices of the spatial firing pattern of pairs of grid cells, i.e. their phase difference.
The functional connectivity takes positive values between cells with nearby phases and 
approaches zero or negative values for larger phase differences.
We also find similar results when, in addition to correlations due to overlapping fields, we
account for correlations due to theta oscillations and head directional inputs.
The inferred connections between neurons can be both negative and positive regardless of whether the 
cells share common spatial firing characteristics, that is, whether they belong to the same modules, or not. 
The mean strength of these inferred connections is close to zero, but the strongest inferred connections 
are found between cells of the same module.
Taken together, our results suggest that grid cells in the same module do indeed form a local network of interconnected
neurons with a functional connectivity that supports a role for attractor dynamics in the generation of the grid pattern.

\newpage 
\section*{Introduction}
Grid cells are neurons in the medial entorhinal cortex (MEC), one synapse away from
the hippocampus, that show a strikingly regular spatial selectivity \cite{Hafting2005}. Each grid cell has several firing 
fields that spread out in a hexagonal pattern, tessellating the environment in which the animal navigates. The 
locations of these firing fields are unaffected by the velocity of the animal, and they persist in 
the absence of external landmarks, suggesting that they make up an intrinsic metric for space 
\cite{Fyhn2004, Hafting2005, Sargolini2006}. These cells were first discovered in rodents 
\cite{Fyhn2004, Hafting2005}, but have recently also been reported in bats \cite{Yartsev2011}, 
monkeys \cite{killian2012map}, and humans \cite{jacobs2013direct}, supporting 
the possibility that grid cells form a part of the neural circuitry underlying the brain's 
internal representation of space in all mammals.

Two main properties of grid cells are their spacing (the shortest distance between two firing fields) and their 
orientation relative to an axis of the environment. Anatomically close grid cells tend to have the same 
orientation and spacing, with spacing increasing along the dorsoventral axis of MEC \cite{Hafting2005, Sargolini2006}. 
This increase is stepwise rather than continuous, such that grid cells can be clustered 
with respect to spacing. These clusters also share other properties, such as orientation, and are therefore referred 
to as modules \cite{Stensola2012}.
A third property of grid cells is their spatial phase, which is defined as the location of the 
grid pattern relative to a reference point in the environment. 
For cells with similar grid pattern, i.e. cells from the same module, one can also measure the difference in spatial 
phase by calculating the shortest distance between firing fields of two cells.
No apparent relationship between the anatomical distance and the difference in spatial phase 
of pairs of neurons has been observed \cite{Hafting2005}.

Since their discovery, grid cells have been under intense investigation, with studies ranging from experimental work to theoretical 
models, in hopes of revealing the underlying network mechanisms behind their coding; see \cite{moser2014network, moser2014grid} 
for recent reviews. In particular, population-wise response properties \cite{Hafting2005,Stensola2012,yoon2013specific} support the idea that the 
formation of grid cells is predominantly a network phenomenon, and that recurrent connectivity in MEC plays an important role.
The main network model of grid cells, the continuous attractor model, would suggest that the hexagonal firing of grid cells
emerges due to specific connectivity patterns between the neurons. In several of these models neurons are considered 
to be arranged in a two-dimensional network according to their phase. Cell pairs beyond a 
certain phase distance inhibit each other,
while those closer to each other are coupled by excitation \cite{McNaughton2006,Fuhs2006,Burak2009}, or less inhibition \cite{Burak2009, Couey2013}, 
as idealized by a `Mexican hat' type of connectivity.

Although connectivity plays important roles in network models of grid cells and in shaping neuronal correlations, 
little has been done to study the correlation structure and 
functional connectivity in the MEC {\it in vivo}, as well as how they change with properties of grid cells, e.g. phase 
separation and theta modulations. In other words, statistical analyses 
of multi-neuronal spike trains of the type routinely performed 
on data recorded from other parts of the nervous system \cite{Truccolo2005,pillow2008spatio,Rebesco2010}, 
is still lacking. Such analyses can shed light on how grid cells encode information at 
the population level and how they interact with each other, providing substance for understanding
the network mechanisms behind the formation of grid cells.

In this paper we aimed at studying the statistical properties of grid cells' multi-neuronal spike trains
by analyzing recordings from two rats while they foraged freely in two-dimensional environments. We therefore first measured the correlations 
between these cells, beyond what is expected from space dependent rate variations, using the same approach as 
\cite{mathis2013multiscale}: we averaged the Pearson correlation coefficients between
firing rates of pairs of neurons during multiple passes
through spatial bins covering the environment. With spatial bins small enough the effect of possible correlations due to rate 
covariations between two cells is removed. These correlations are referred to as {\it noise correlations}.
We found that these correlations decay as the phase
difference between cell pairs increases. This is consistent with previous analyses of pairs of grid cells recorded on
a linear track \cite{mathis2013multiscale}. Second, we fit a statistical model that assumes 
a pairwise maximum entropy distribution 
over the spikes generated in a time bin, given the spike pattern in the previous time bin and external covariates 
also referred to in the text as {\it external fields}.
This model is known in the statistical physics community as the kinetic Ising model and belongs to the class of generalized linear 
models (GLMs) \cite{nelder1972generalized} 
with short time memory kernels. We considered an extensive list of external covariates 
known to modulate the firing of grid cells to explain the covariations in firing rates of neurons, ranging from spatially and temporally 
constant input, 
to spatial fields formed as the sum of Gaussian basis functions, as well as fields for speed, theta oscillations,
and head and running directions.
We evaluated the explanatory power of these models by comparing their likelihood values and found that speed, 
head direction and running direction had 
little power in explaining the data, while theta oscillation phase and pairwise 
couplings had more explanatory power. Although there were variations in terms of the relative strength of the 
couplings depending on the assumptions about the external fields, 
we consistently found that the inferred connections maintained a pattern that supports the attractor network hypothesis: 
cells with nearby phases tend to excite each other while those further apart inhibit each other. 
We also found that the strongest connections were among cells within the same module, that the connections were both 
negative and positive, and that none of our conclusions were sensitive to data limitations. 

\section*{Results}
We analyzed two data sets with simultaneously 
recorded grid cells, one with a total of 65 cells, of which 27 were grid cells (referred to as data set 1), the other with 8 grid cells (data set 2).
As mentioned, grid cells are known to cluster according to the spacing and orientation of their spatial fields,
with cells with similar spacing making distinct functional modules that react in unison to external manipulations of the environment
as quasi-independent populations \cite{Stensola2012}. In data set 1, all but 5 of the grid cells were easily identified into 
three distinct modules (see Material and Methods). In data set 2, all 8 cells belonged to the same module.

\subsection*{Noise correlations}
\begin{figure}[h!]
\begin{center}
\includegraphics[width=48mm]{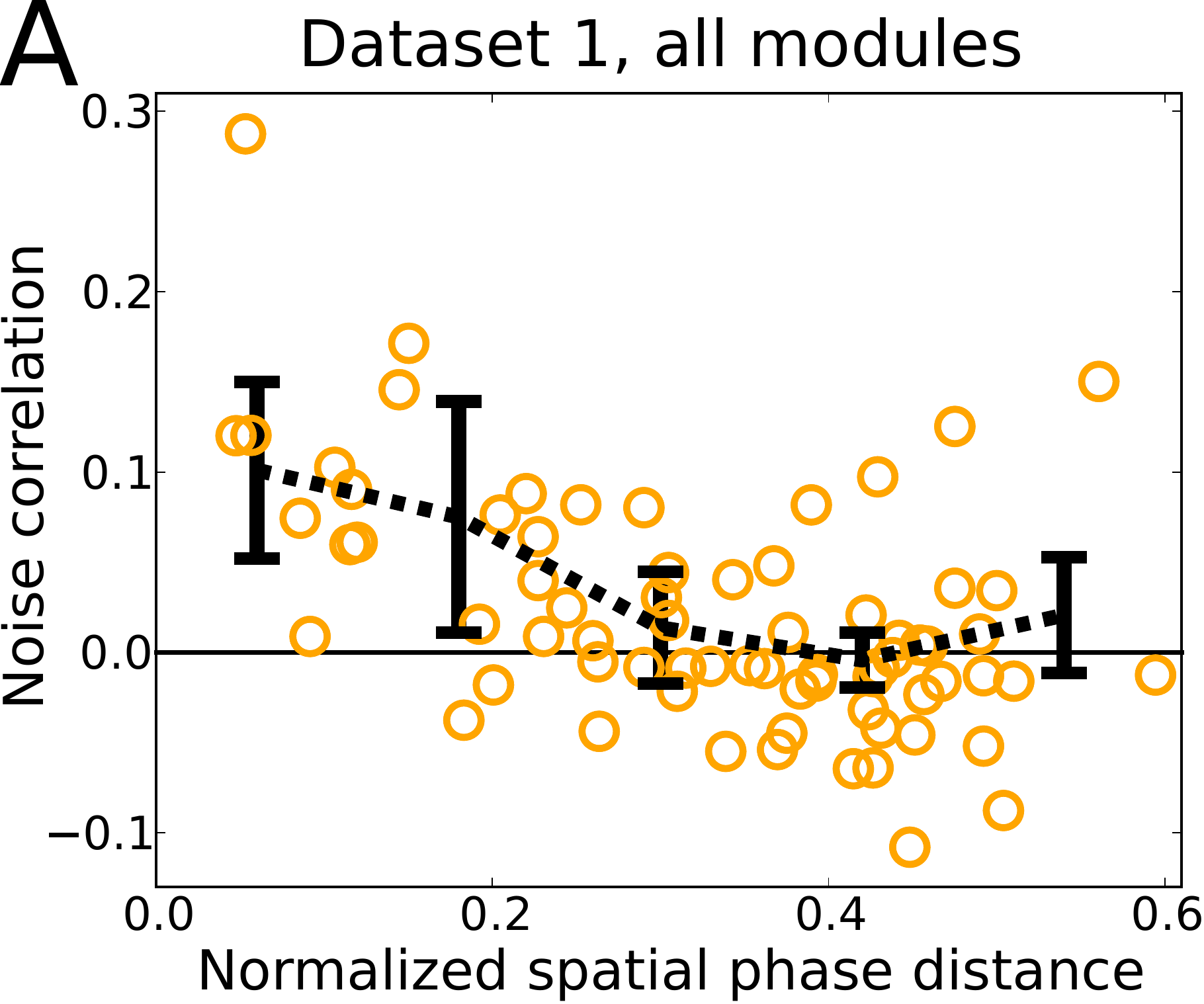}  
\includegraphics[width=48mm]{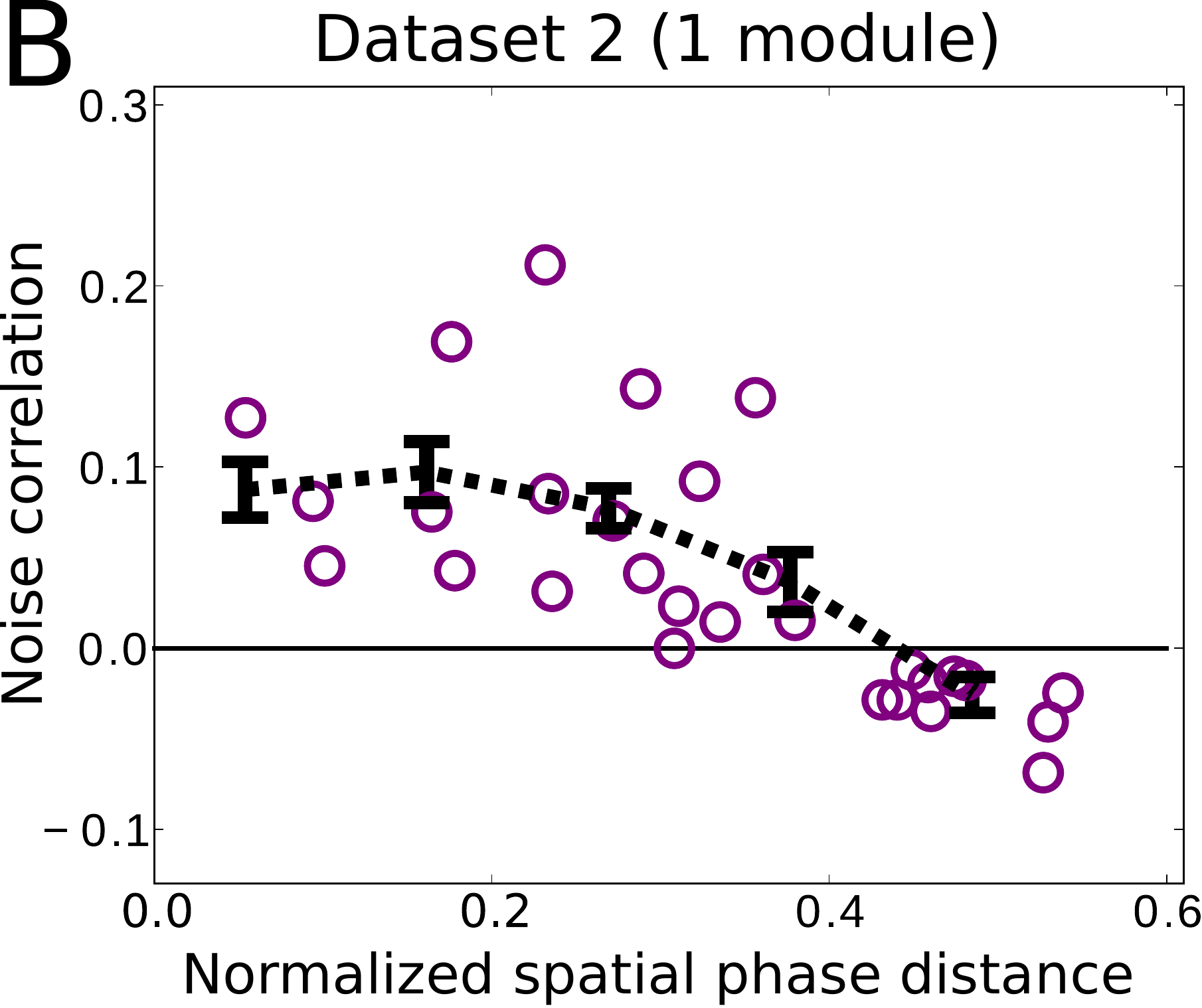}\\ \vspace{0.5cm}
\includegraphics[width=48mm]{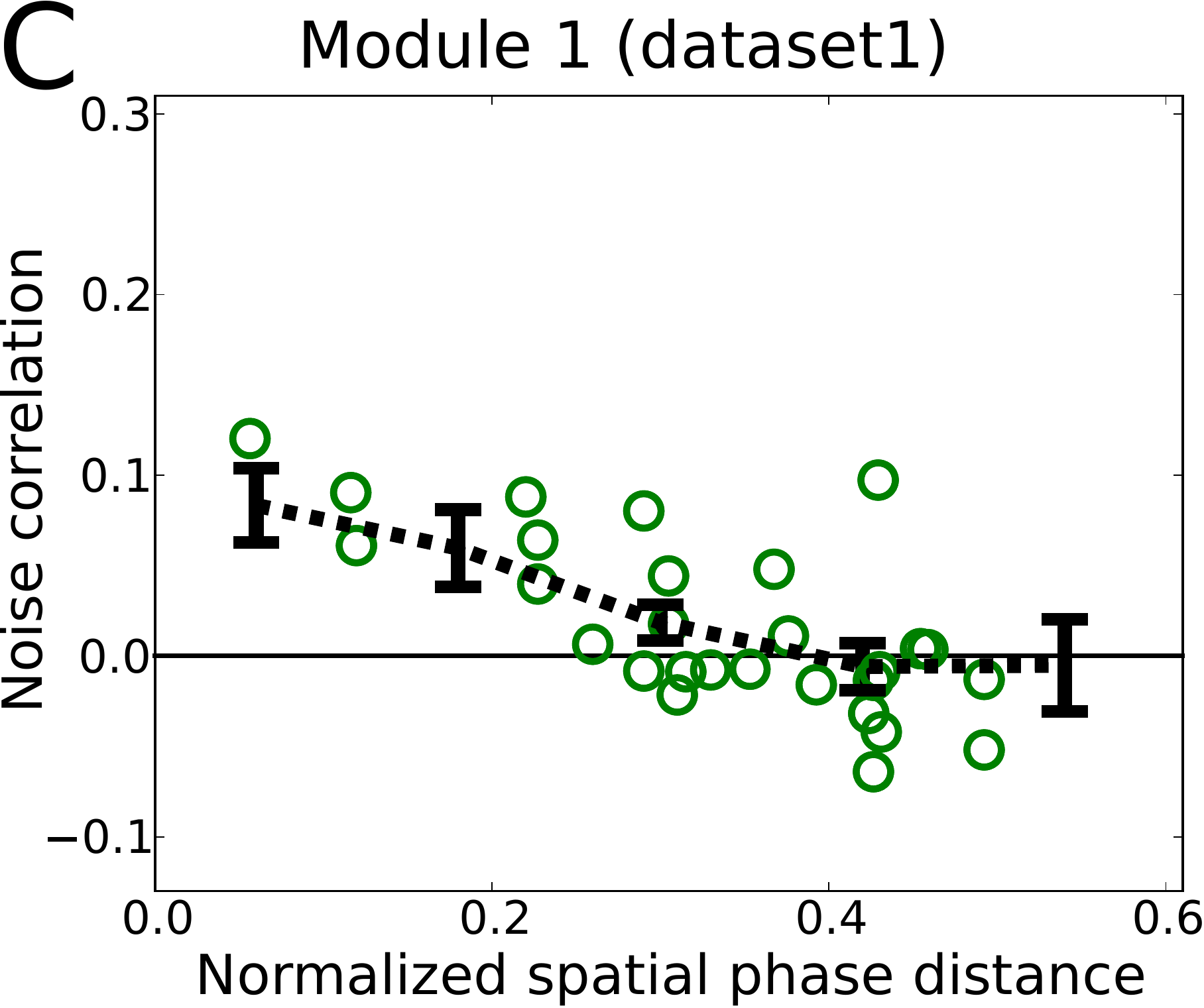}
\includegraphics[width=48mm]{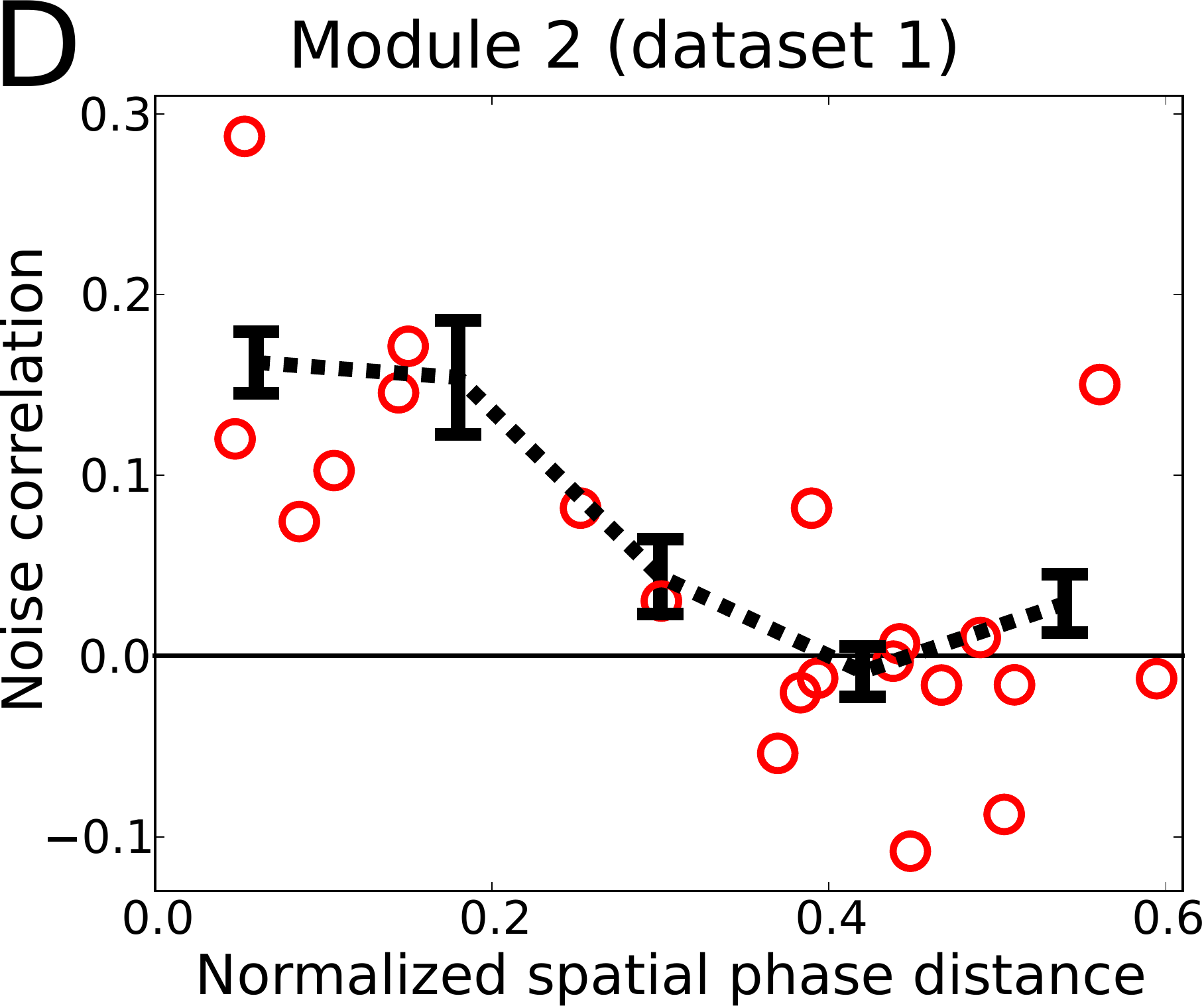}
\includegraphics[width=48mm]{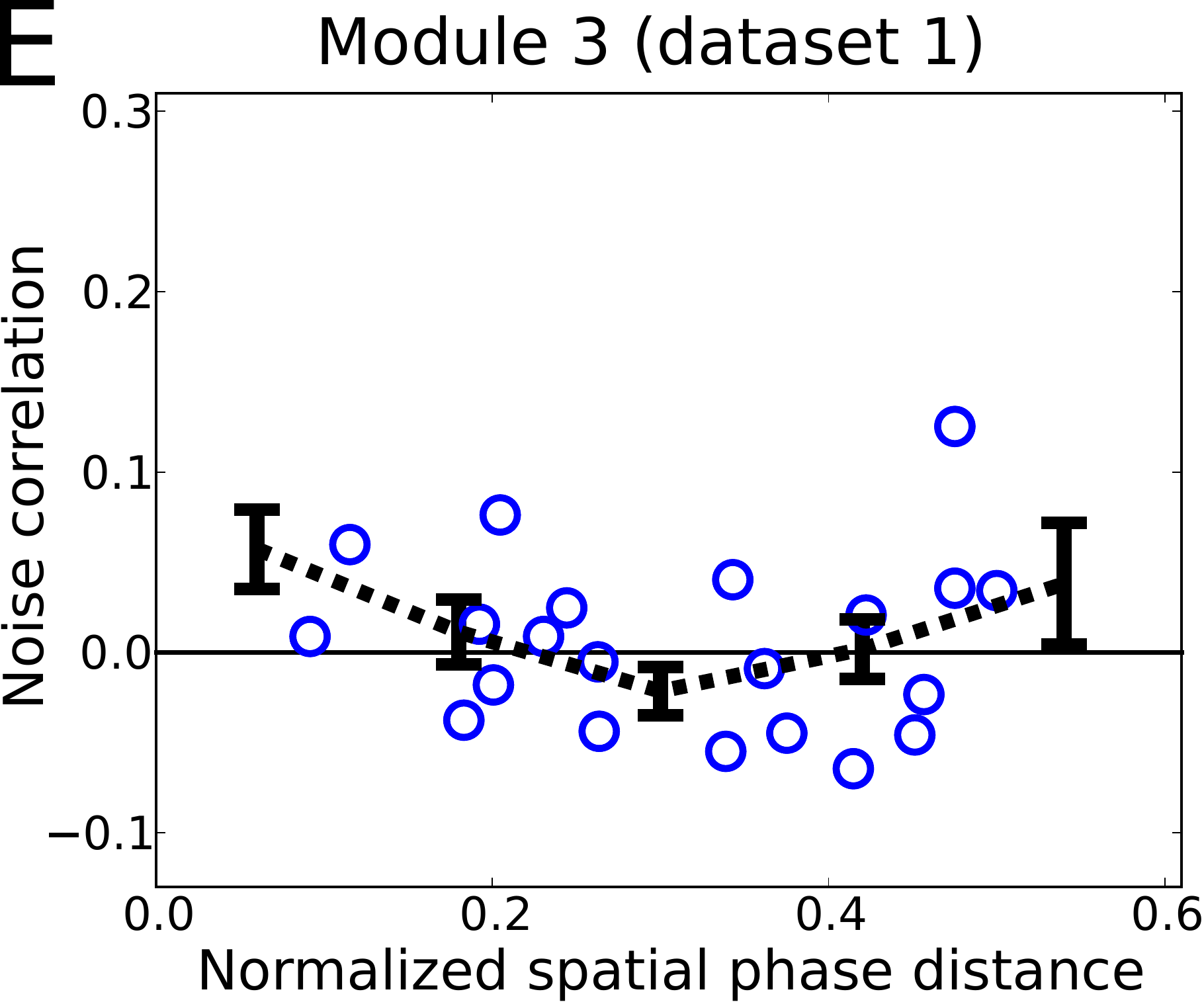}
\end{center}
\caption{{\bf Noise correlations versus phase distance.} \small{(A) shows all three modules of data set 1 combined, 
while {(B) shows the one module of data set 2. For the smaller modules of data set 1 (C, D)} the noise correlations are positive for 
small phase differences while they approach zero for larger phase separations. 
No significant pattern can be observed for the cells from the largest module of data set 1 (E). 
The distance in phase was normalized by the average spacing of the spatial fields in each module.
In each plot, the circles represent the inferred values using the full data length.
The noise correlations were calculated by binning the environment into $7.5\times7.5$ cm spatial bins.
The black lines show the average values of the correlations calculated from 20 
random partitions (see Material and Methods) of the data. The error bars are the standard deviation of the mean values over these
20 random partitions.  Note that the normalized maximal phase distance occurs at the minimum overlap between the two commonly oriented 
hexagonal patterns and is $0.5/\cos(30) \approx 0.6$.}}
\label{FigCorrelationvsPhaseDist}
\end{figure}

To calculate correlations between pairs of grid cells, beyond what is expected from spatial rate covariations,
we binned the spike data into 1 ms intervals and smoothed the firing rates with a 20 ms Gaussian filter. 
The trajectory of the animal was then binned spatially by dividing the environment into a number of $N\times N$ square boxes, using different values of $N=2, 3, 4, 5, 10, 15, 20, 40, 75$. 
Noise correlations, $C_{ij}$, between cells $i$ and $j$ were then determined as the mean of the Pearson correlation coefficients, $\rho$, 
calculated over the trajectories through each spatial bin (see Material and Methods).
As shown in Fig.\ \ref{FigCorrelationvsPhaseDist}, in the case of dividing the environment into $20\times 20$ spatial bins, 
we found noise correlation values close to zero, or slightly negative, for cells with non-overlapping spatial fields. 
On the other hand, cell pairs close in phase distance showed positive noise correlation values that increased for cells closer to each other in 
phase; see Fig.\ \ref{FigCorrelationvsPhaseDist}A and B. The slope ($\hat{\beta}$) and intercept 
($\hat{\alpha}$) of a linear regression line (not shown) are $\hat{\beta}$ 
= $-$0.22 and $\hat{\alpha}$ = 0.09 for data set 1, and $\hat{\beta}$=$-$0.25 and $\hat{\alpha}$ 
= 0.11 for data set 2, all significantly different from 0 (t-test, P$<$0.001).
 
Since data set 1 included neurons from 3 separate modules, we also studied the dependence of the noise correlations on the phase difference between 
cells for each three modules separately. 
Except for the module with the largest field spacing (Fig.\ \ref{FigCorrelationvsPhaseDist}E), where the phase dependence was weak (intercept and slope of
linear regression not significantly different from 0 (t-test, P$>$0.7)), the modules 
showed a significant pattern similar to that of all modules pooled together shown in Fig.\ \ref{FigCorrelationvsPhaseDist}A (intercept and slope of linear 
regression significantly different from 0 (t-test, P$<$0.001)). Similar results were found when other spatial bin sizes were used.  This extends the results of 
\cite{mathis2013multiscale} to two dimensions and also shows the variations in the phase dependence of 
the correlations to the module size.

Good empirical estimates of the noise correlations, as defined above, require that the rat makes enough passes through each spatial bin during the recording session. 
This means that the bins cannot be too small, otherwise there 
would be very few visits to most of the bins, and some of the bins may never be visited at all.
On the other hand, if the bins were too big, the variations in rate from one pass through the bin to another 
would be be too large and, therefore, $C_{ij}$ would not exclude the rate covariations. 
We, therefore, looked at how consistent our estimates of the correlations were as a function of the spatial bin size by 
calculating the Pearson correlation coefficient between the correlations measured, using a random half of the visits to each spatial bin 
with those measured from the other half (see Material and Methods). 
The most stable estimate was with $20$ bins per side of the box (or 7.5 cm), which is what we have used in 
Fig.\ \ref{FigCorrelationvsPhaseDist}. In this best case scenario, for data set 1, the Pearson correlation coefficient is $0.56$ for the 
full data, with both halves of the data in all 20 sets of random halves still 
demonstrating the phase dependent pattern shown in Fig.\ \ref{FigCorrelationvsPhaseDist}. 
Cells with nearby grid patterns had stronger positive correlations, while those further apart 
in phase demonstrated a slightly negative, or no correlations (the slope and 
intercept of the linear regression lines were all significantly different from 0 (t-test, P$<$0.03)). 
This was also the case for the 20 random halves of data set 2.

The pairwise correlation analysis done here is a good first step, however, it suffers from a number of shortcomings. First of all, it is really 
a pairwise measure, which excludes the interactions with other neurons, and thus a perceived correlation between two cells 
might really be explained by the presence of a third neuron or external covariates. Second, although 
we take into account spatial covariations in rate, there is no systematic way of evaluating how 
much other covariates, such as theta oscillations or head direction, contribute to the 
correlations between cells. 
Given the fact that grid cells are known to covary with these, it is important to evaluate 
their influence when analyzing correlations between grid cells. 
While pairwise correlation analysis suffers from these problems, they can be addressed,
to a large extent, using statistical models of the GLM type. This is what we will do in the rest of the paper.  
 
\subsection*{Functional couplings and the effect of external covariates}
As a statistical model, we considered the simplest maximum entropy model to include both asymmetric couplings and time 
varying external input: the kinetic Ising model. 
The activity of the cells was binned in 10 ms bins, and a binary variable $S_i(t)$ was associated to each neuron in each bin,
which would be equal to +1/-1 denoting the presence/absence of
spikes emitted by neuron $i$ within time bin $t$. Letting the state of each neuron at time $t$ depend on the state 
of the population in the previous time step $t-1$ and some covariates, independent of 
the state of the system, the maximum entropy distribution over the state $S_i(t)$ of neuron $i$
at time $t$ is \cite{jaynes1982rationale}
\begin{eqnarray}
\label{eq:kineticIsingPdist}
   P(S_i(t) | \{{}\mathbf{S}(t-1)\}{}) = \frac{\exp[S_i(t)H_i(t-1)]}{2 \cosh[H_i(t-1)]} , \\
  H_i(t-1) = h_{i}(t-1) + \sum\nolimits_{j}{J_{ij}S_j(t-1)}
\end{eqnarray}
where $J_{ij}$ would be identified as the functional coupling from neuron $j$ to neuron $i$,  and $h_i(t)$ as the time varying
covariate which in statistical physics terminology is called an external field. As mentioned
in the introduction, Eq.\ \ref{eq:kineticIsingPdist} defines a GLM, where in each time bin, mostly only one or zero spikes per bin are observed and the 
interaction kernel extends one time step in the past.
With binary states and only one time step kernels, 
this model represents the simplest possible model capable of capturing functional connectivity
from neural data, which is convenient given the finite time in which the neural recordings were taken.
This model should not be confused with the maximum entropy equilibrium models (equilibrium Ising model \cite{schneidman2006weak,shlens2006structure}), 
which assume symmetric couplings and are not related to the GLMs.

Given Eq.\ \ref{eq:kineticIsingPdist}, we asked what values of the parameters 
$h_{i}(t)$ and $J_{ij}$ are the most likely to generate the observed data. 
 Both exact and fast approximate algorithms for solving the inverse kinetic Ising model 
have been developed \cite{Roudi2011meanfield} similar to other GLM models \cite{nelder1972generalized,Truccolo2005,pillow2008spatio}. 
The exact solution is found by maximizing the log-likelihood function
\begin{equation}
L[\textbf{S},\textbf{J},\textbf{h}] = \sum\nolimits_{it}{\big[S_i(t+1)H_i(t) - \text{log 2 cosh }H_i(t)\big]}
\label{eq:likelihood}
\end{equation}
with respect to $h_{i}(t)$ and $J_{ij}$. The term `exact' is used here in the sense that if data is generated by a kinetic Ising model, 
this learning algorithm would recover the parameters exactly in the limit of infinite data. The log-likelihood is the logarithm of 
the probability of observing the data at hand given that it was generated from the model, and thus measures 
how well the model explains the statistics in the observed data. In our analysis 
we have used the natural logarithm.
\begin{figure}[h!]
\begin{center}
\begin{tabular}{ccc}
  \includegraphics[width=45mm]{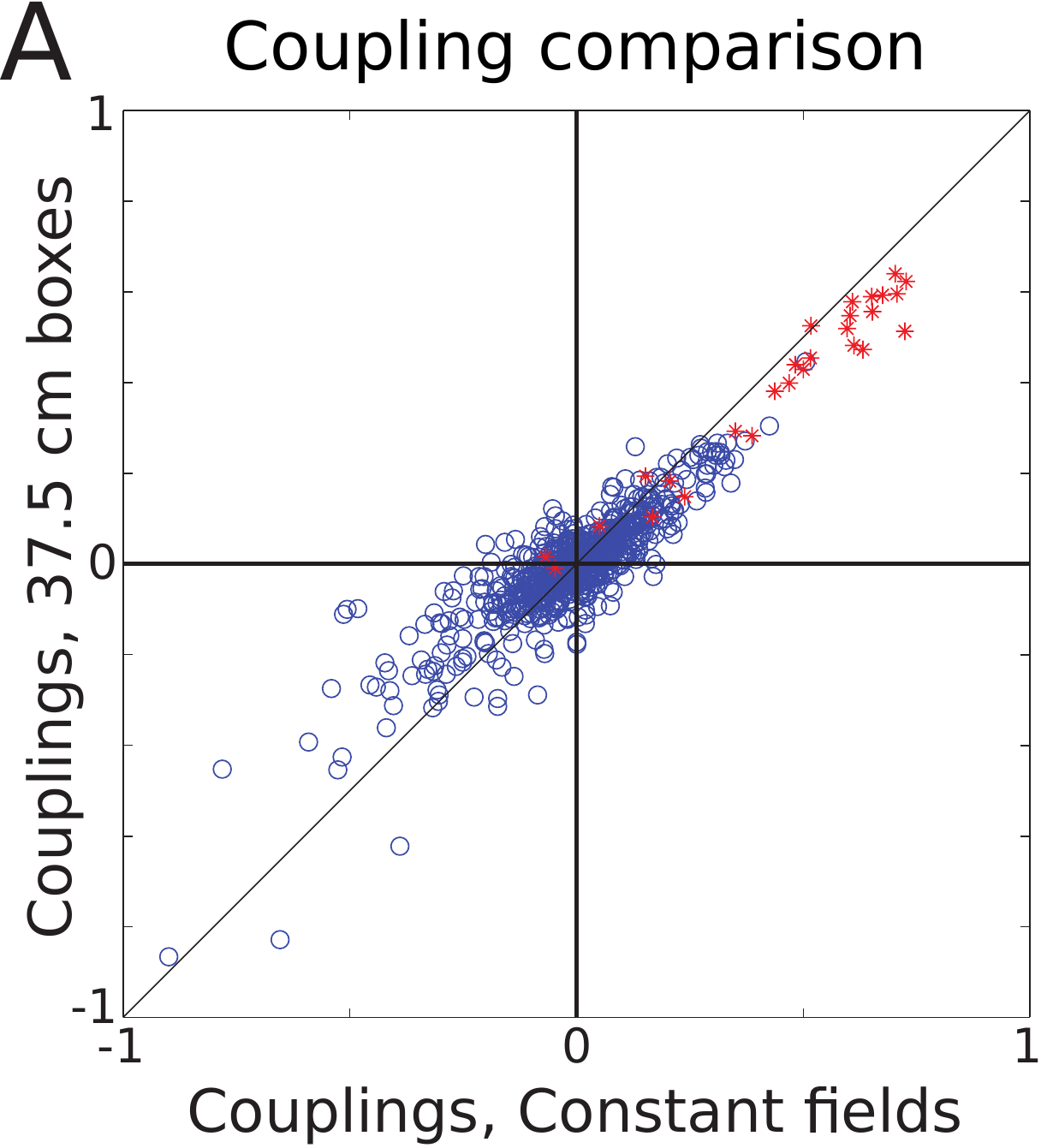} &
    \includegraphics[width=45mm]{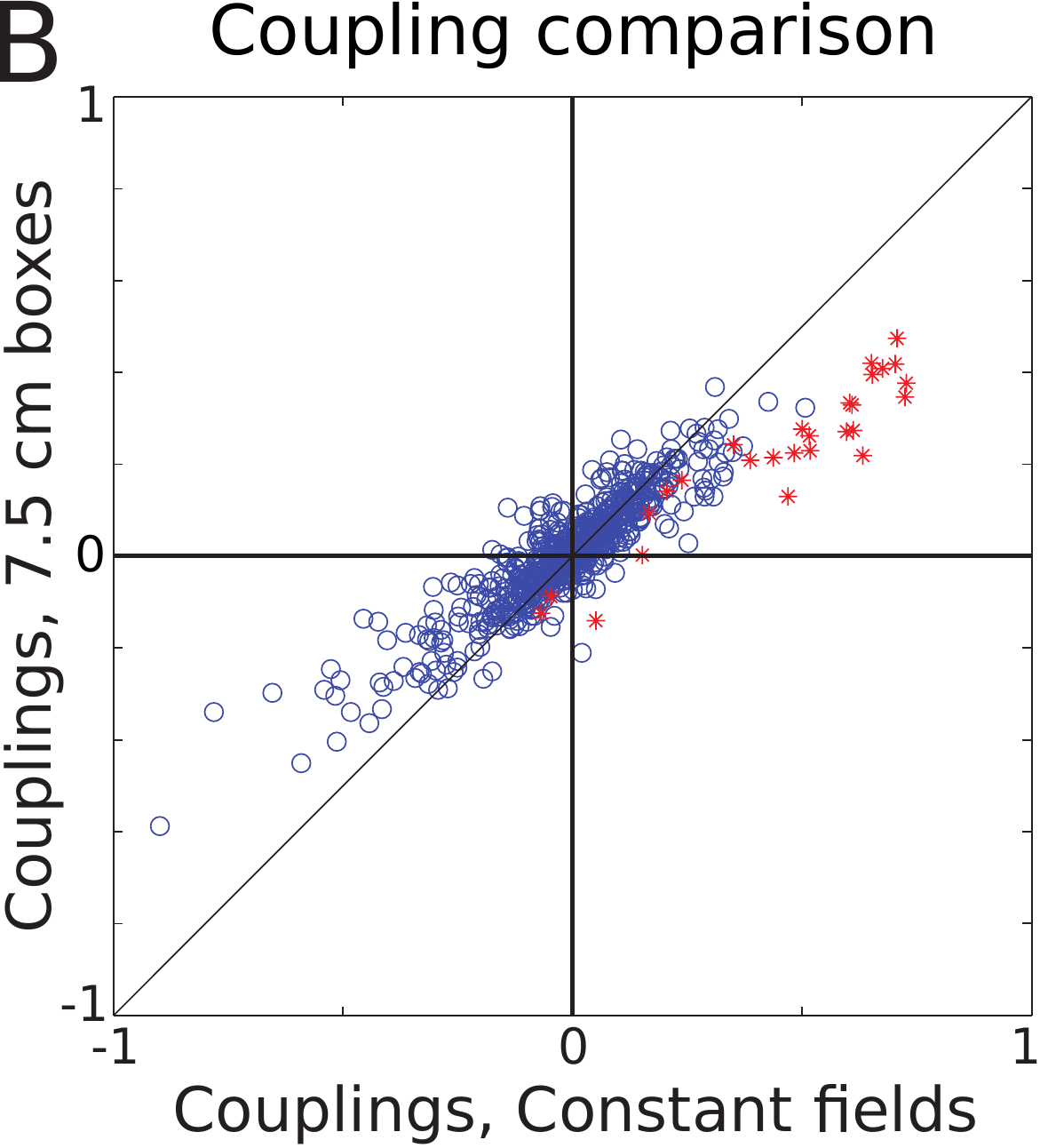} &
  \includegraphics[width=45mm]{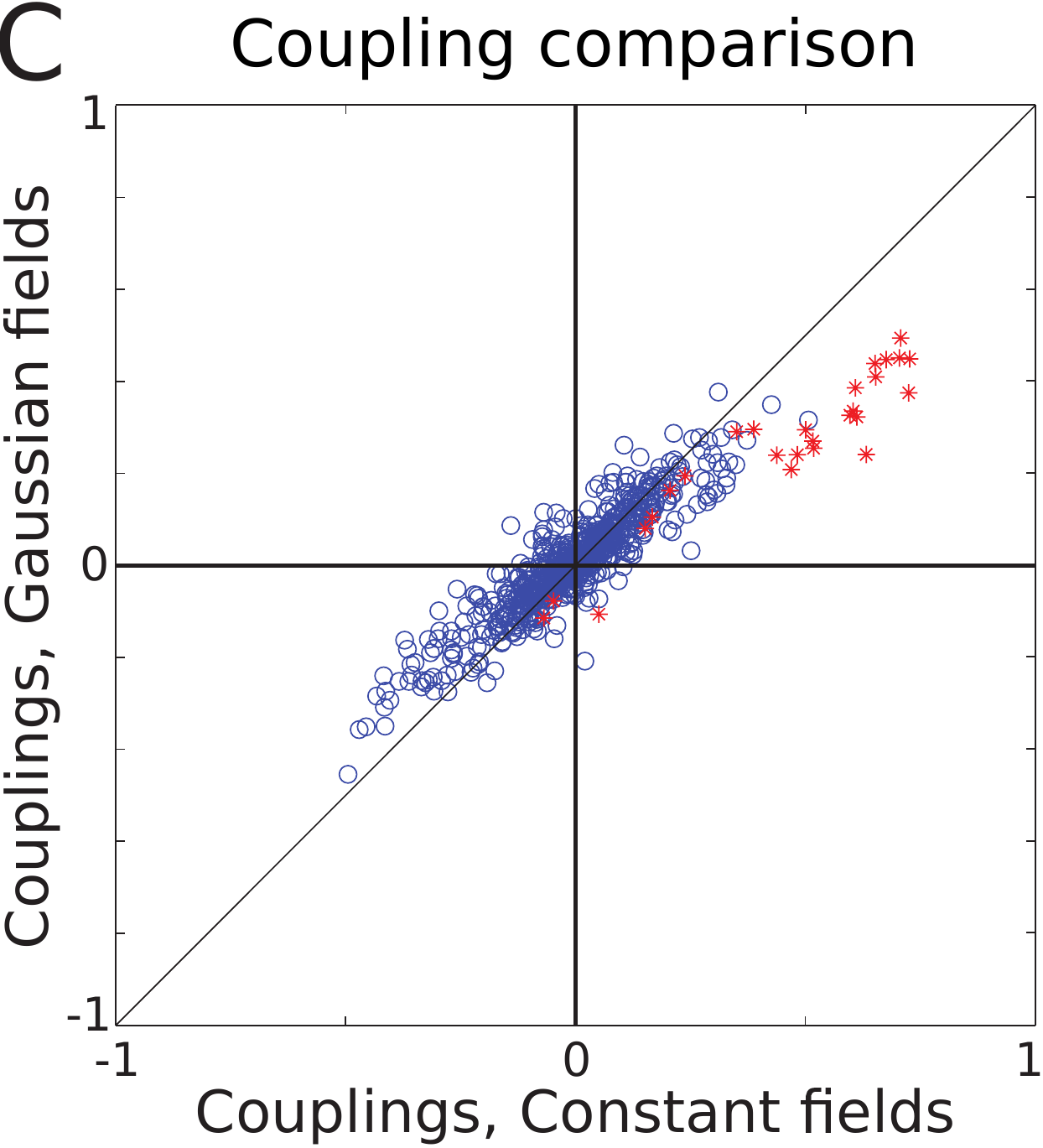} \\
  \end{tabular}
\end{center}
\caption{{\bf The couplings of the kinetic Ising model.} \small{We considered different forms of spatial external input 
to the neurons, boxes of length 37.5 cm (A), 7.5 cm (B) and fields formed as a weighted sum of Gaussian basis functions (C) for data set 1.
For each case, we compared the resulting couplings to that of a model with spatially and 
temporally constant fields. The effect of input with spatial variation is to slightly weaken the couplings.
Pearson correlation coefficient (PCC) was calculated for all the couplings together (All), as well as for just the self-couplings (SC) shown by red stars, and the non-self-couplings 
(NonSC) shown by blue circles. The corresponding values are A: PCC, All = 0.91, PCC, SC = 0.98, PCC, NonSC = 0.86. B: PCC, All = 0.91, PCC, SC = 0.94, 
PCC, NonSC = 0.90.
C: PCC, All = 0.92, PCC, SC = 0.94, PCC, NonSC = 0.91.}}
\label{FigConsistCwithH}
\end{figure}

An important issue in dealing with a model of this type is choosing the external field. 
In the absence of couplings, the external field, $h_i(t)$, can explain the variations in the firing rate as 
the rat navigates in space. Ideally, the external fields can be inferred by binning the environment into small 
spatial bins, assuming that the external field in each bin takes a constant value for each neuron. 
If the rat passes through each bin many times, the external 
field in each bin can be reliably estimated. However, during a recording period, and as described above, the 
requirement of passing through small spatial bins many times is rarely satisfied.  

Alternatively, the spatial input could arise as the sum of two-dimensional Gaussian basis functions with
the basis set spanning the environment. By inferring the parameters of a linear combination of Gaussian basis functions (see Material and Methods for details), an accurate 
representation of the spatial field can be found, even with a reduced amount of data, as shown in the following.

\begin{figure}[h!]
\begin{center}
\begin{tabular}{ccc}
  \includegraphics[width=45mm]{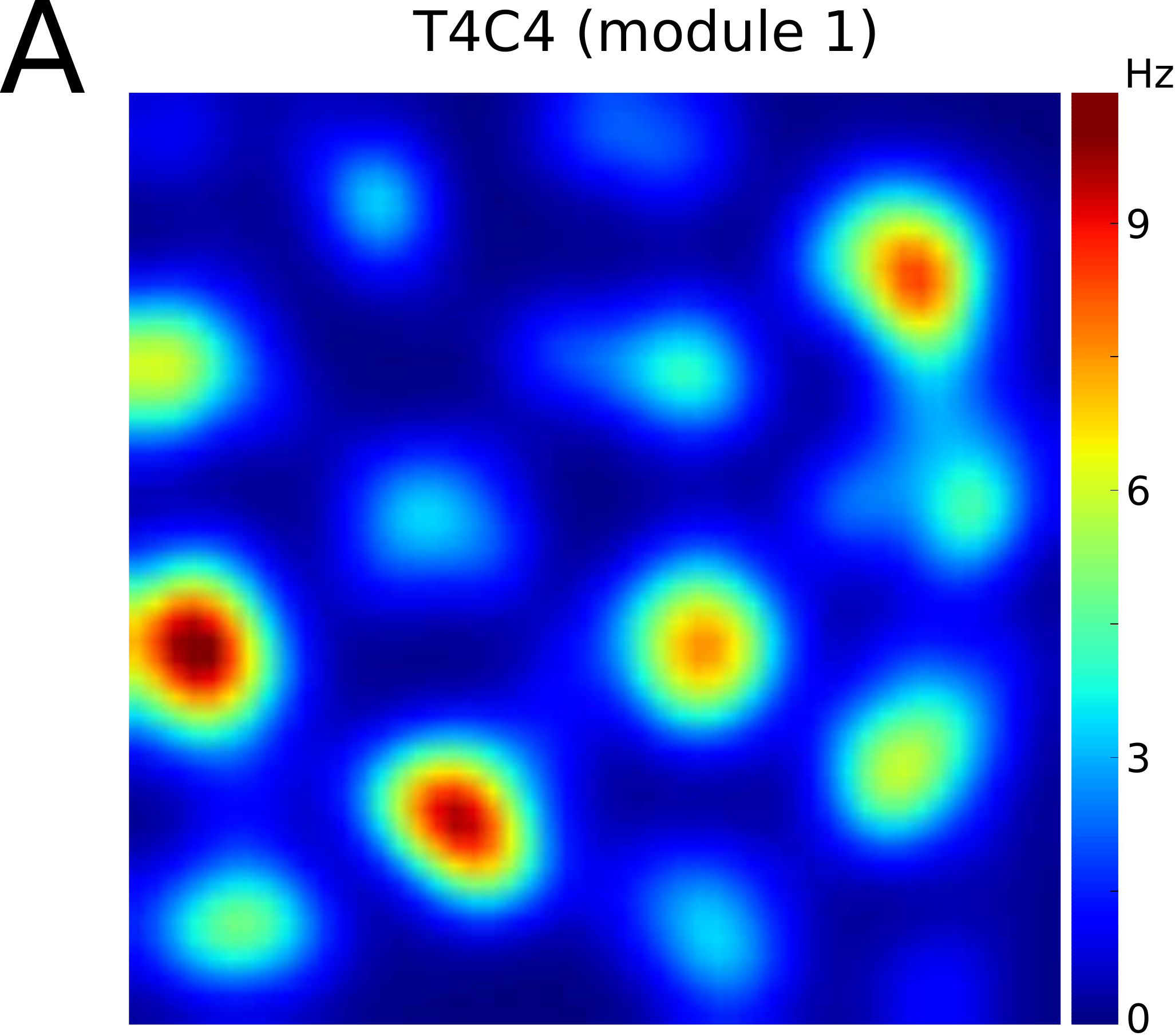} &
    \includegraphics[width=40mm]{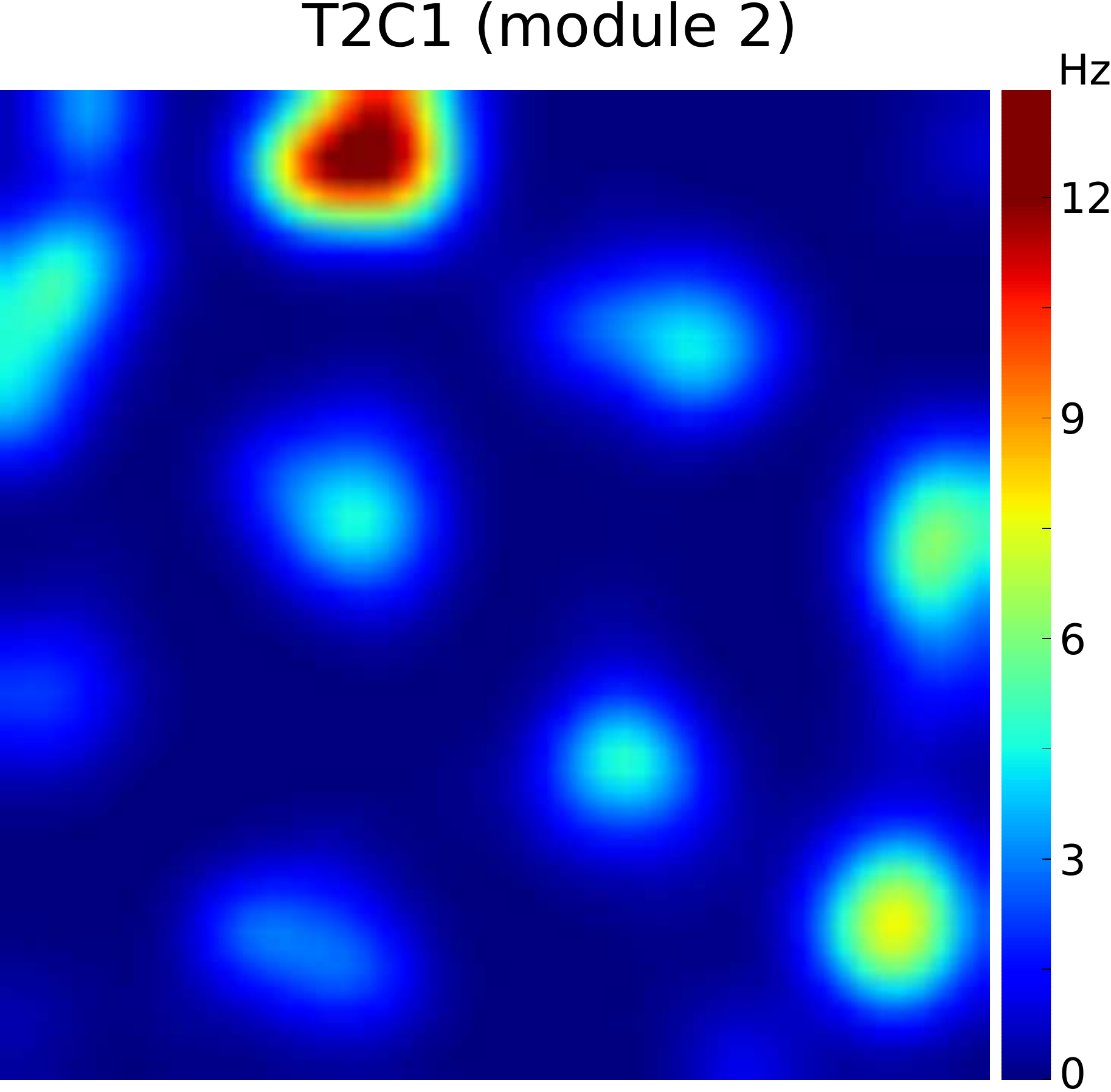} &
  \includegraphics[width=40mm]{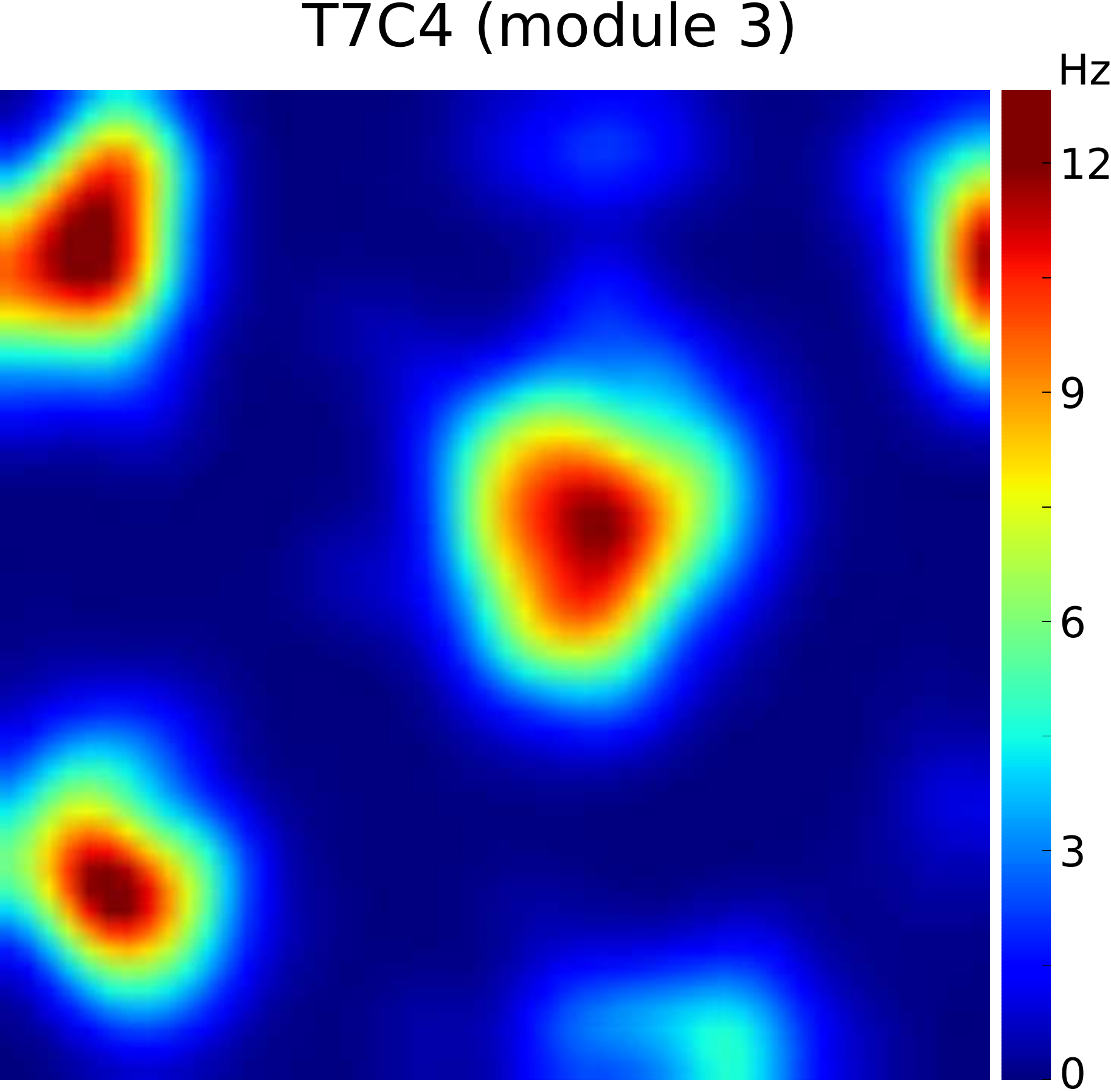} \\
  \includegraphics[width=44.5mm]{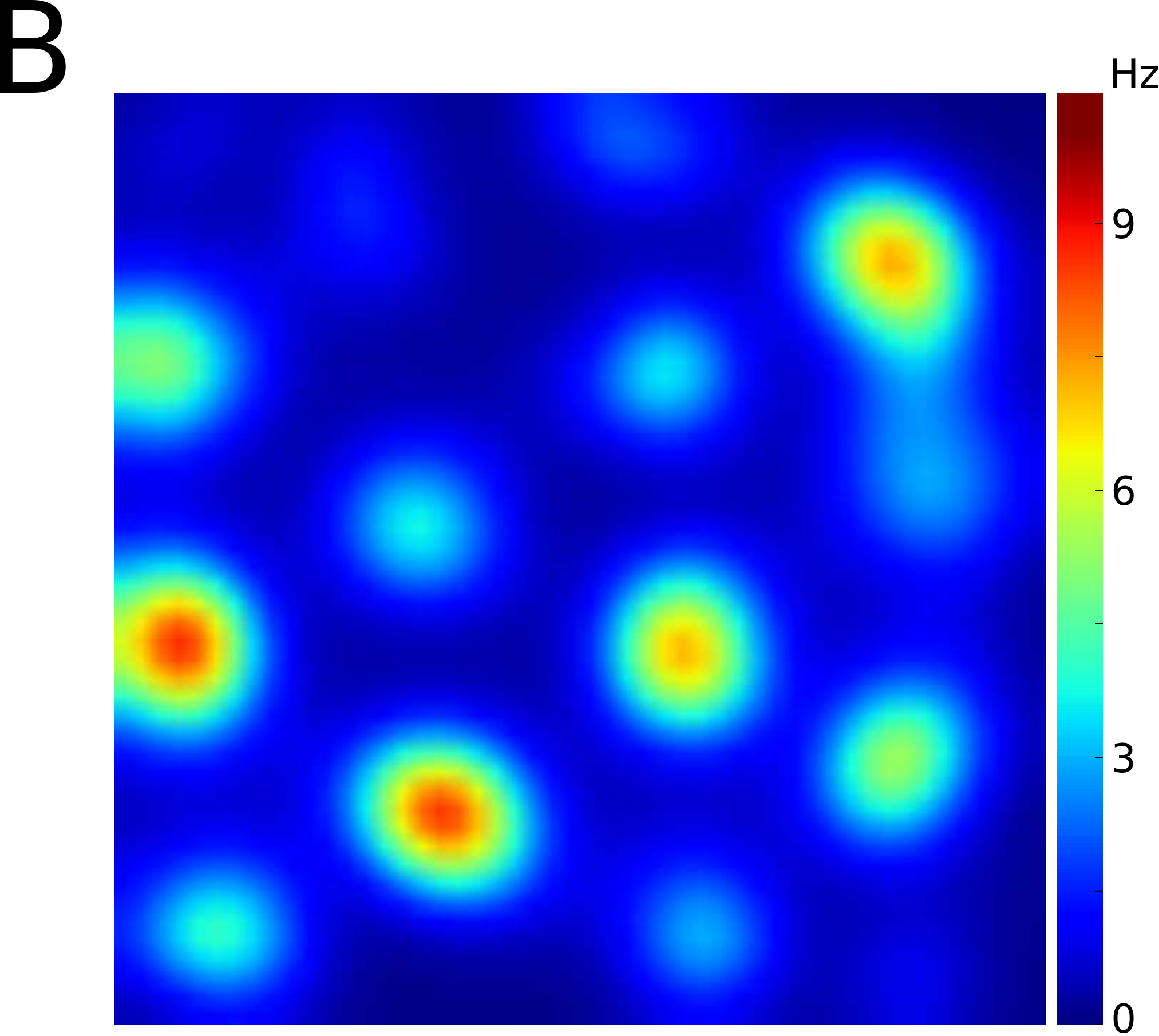} &
    \includegraphics[width=40mm]{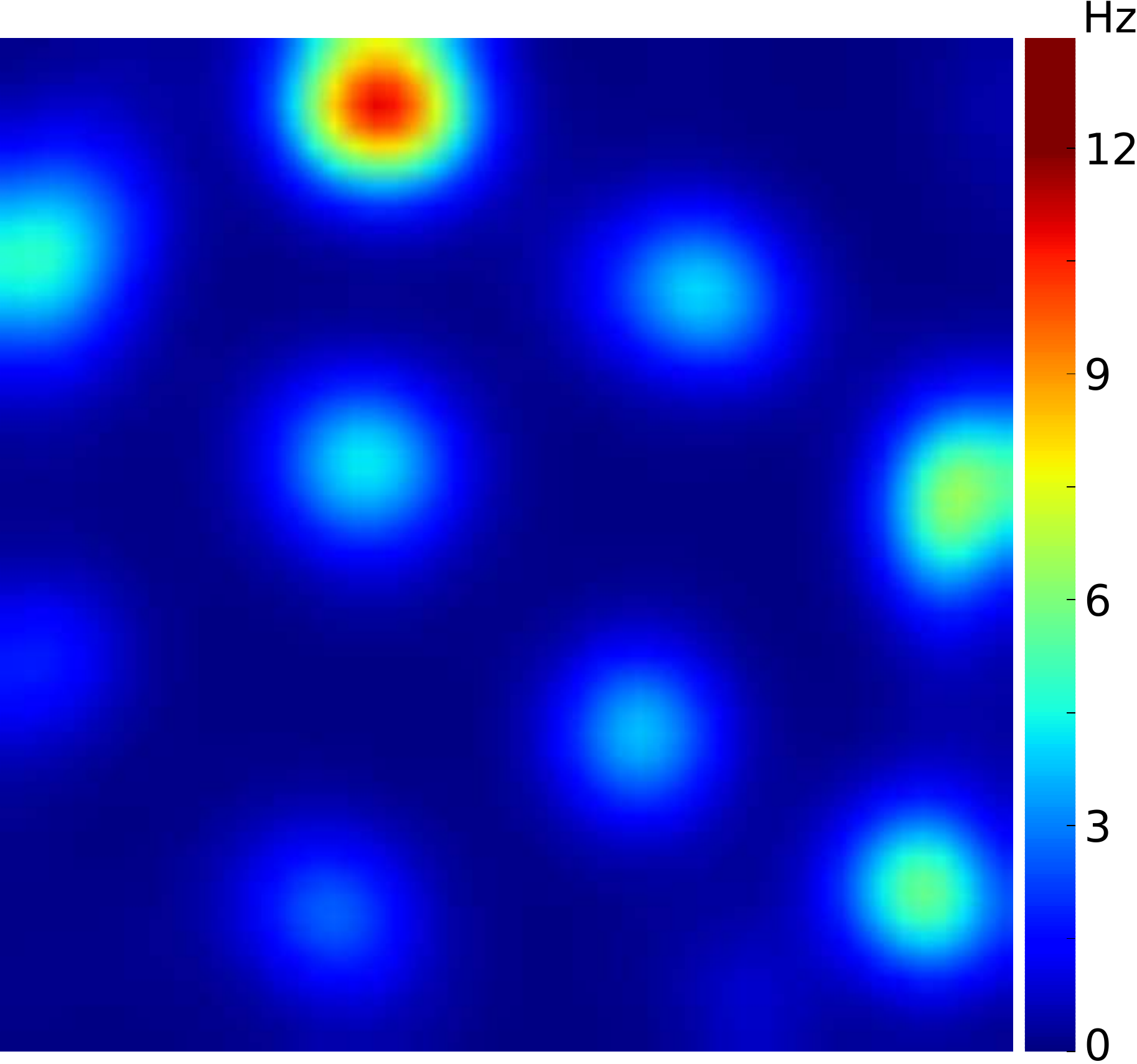} &
  \includegraphics[width=40mm]{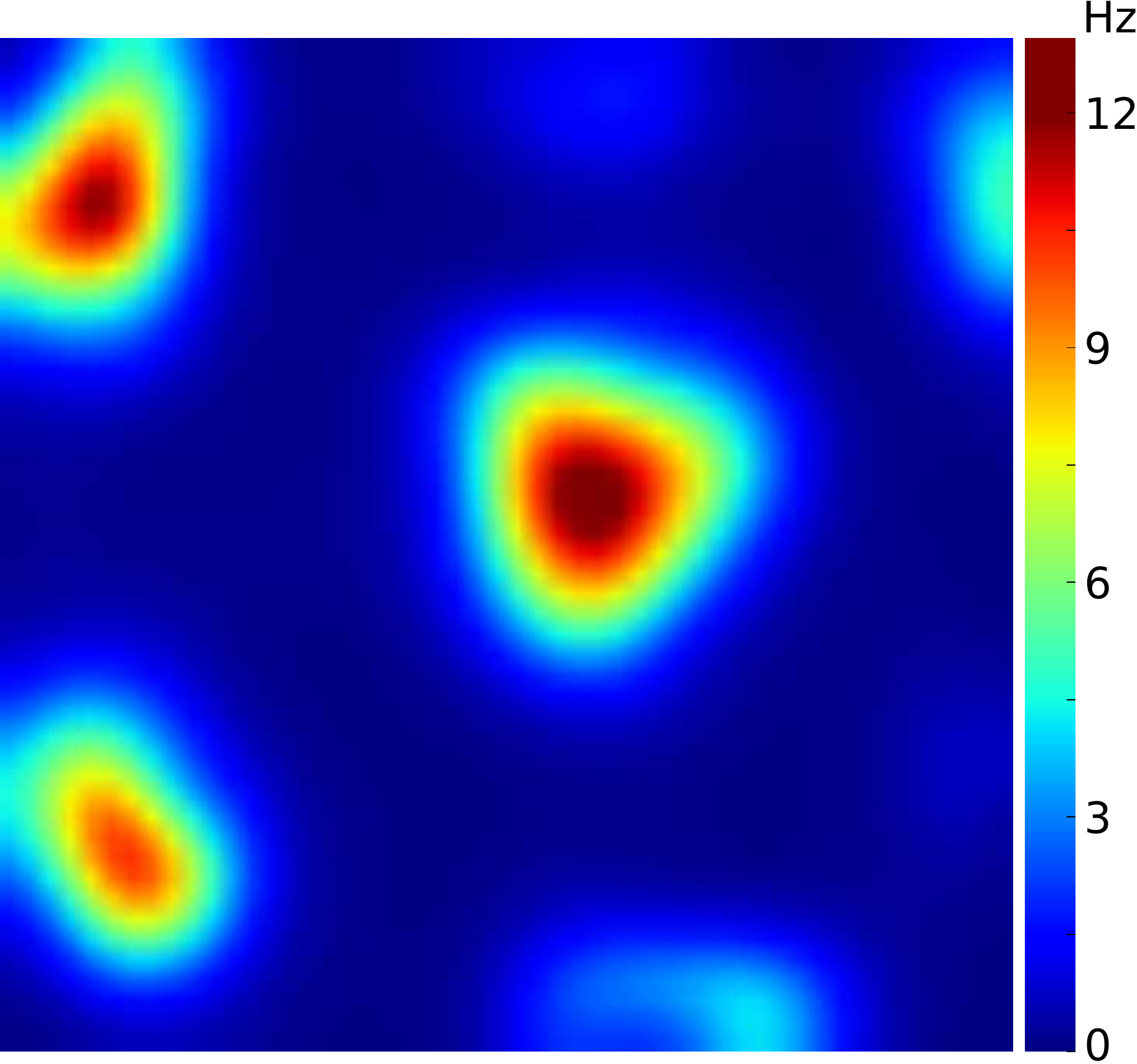} 
  \end{tabular}
\end{center}
\caption{{\bf Grid cell spatial firing rate map.} \small{The smoothed rate maps from the original
spike data (panel A) and synthetic spike data (panel B). Three example grid cells from the three different modules identified in data set 1 are shown here: 
left column, module 1 (T4C4 is cell identity - tetrode 4 cell 4), middle column, module 2, and right column, module 3. The synthetic data (panel B) was generated using Eq.\ \ref{eq:kineticIsingPdist} 
(with $H_i(t)$
determined by the inferred values for the Gaussian basis functions plus a constant field) and the trajectory of the rat. The rate maps in both panel 
A and B were generated by first binning the spike data into 3 cm spatial bins, for which the mean rate was calculated and then smoothed using a 
Gaussian filter (standard deviation = 2 bins).}}
\label{FigSpatialFields}
\end{figure}

Focusing on data set 1, which had the most cells, we first inferred couplings, assuming that each neuron receives an external field 
which is constant across time and space, $h_i(t) = h_i$. Next, we studied how the inferred couplings were affected by increasing the spatial 
resolution of the external fields, $h_i(t)$, to account for the spatial variation in firing rate by dividing the environment into spatial 
bins, considering the cases of bins of size 37.5 cm and then bins of size 7.5 cm, assigning one external field per box to each cell. 
We also considered external fields in the form of a sum of Gaussian basis functions.
Fig.\ \ref{FigConsistCwithH} shows the resulting couplings, plotted against 
couplings found in the model that assumed spatially and temporally constant external input, $h_i$, 
for each neuron. As can be seen, increasing the resolution 
of the external fields made the couplings weaker but not inconsistent with the constant field case, even in the case of Gaussian fields,
where the spatial rate maps were well captured by the model, as shown in Fig.\ \ref{FigSpatialFields}. In this case, 
there was a significant weakening of the couplings (the estimated variance of the Gaussian field model couplings ($S^2_{\text{Gauss}}$) was
significantly smaller than that of the constant field model ($S^2_{\text{constant}}$), 
(F-test for equal variances, P$<$0.001)).
In each of the models, the total external fields were negative and often strong, 
as one would expect for data sets with low firing rates (mean firing rate 2.4 Hz).
   
\begin{figure}[h!]
\begin{center}
\begin{tabular}{cc}
\includegraphics[width=48mm]{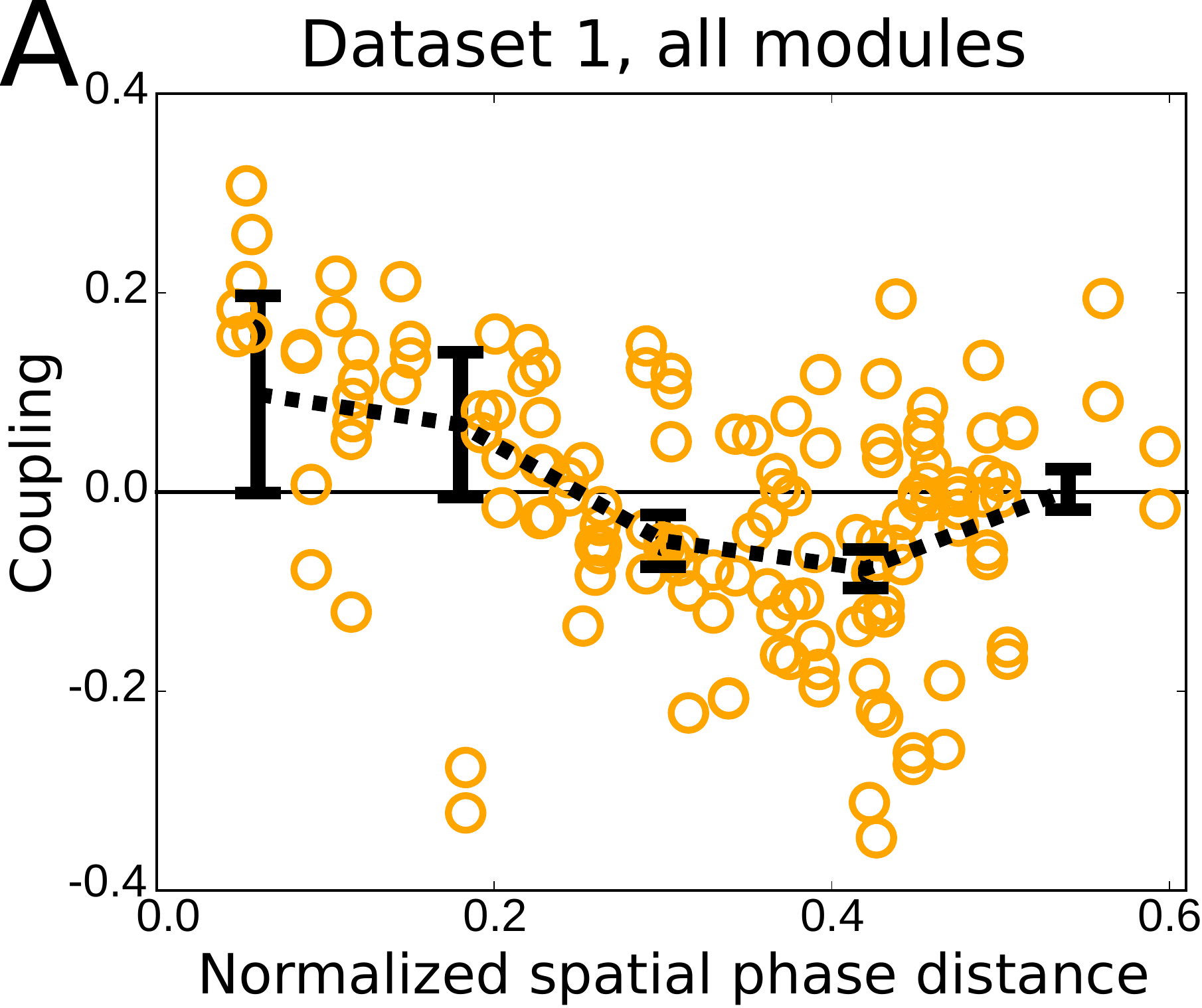} 
\includegraphics[width=48mm]{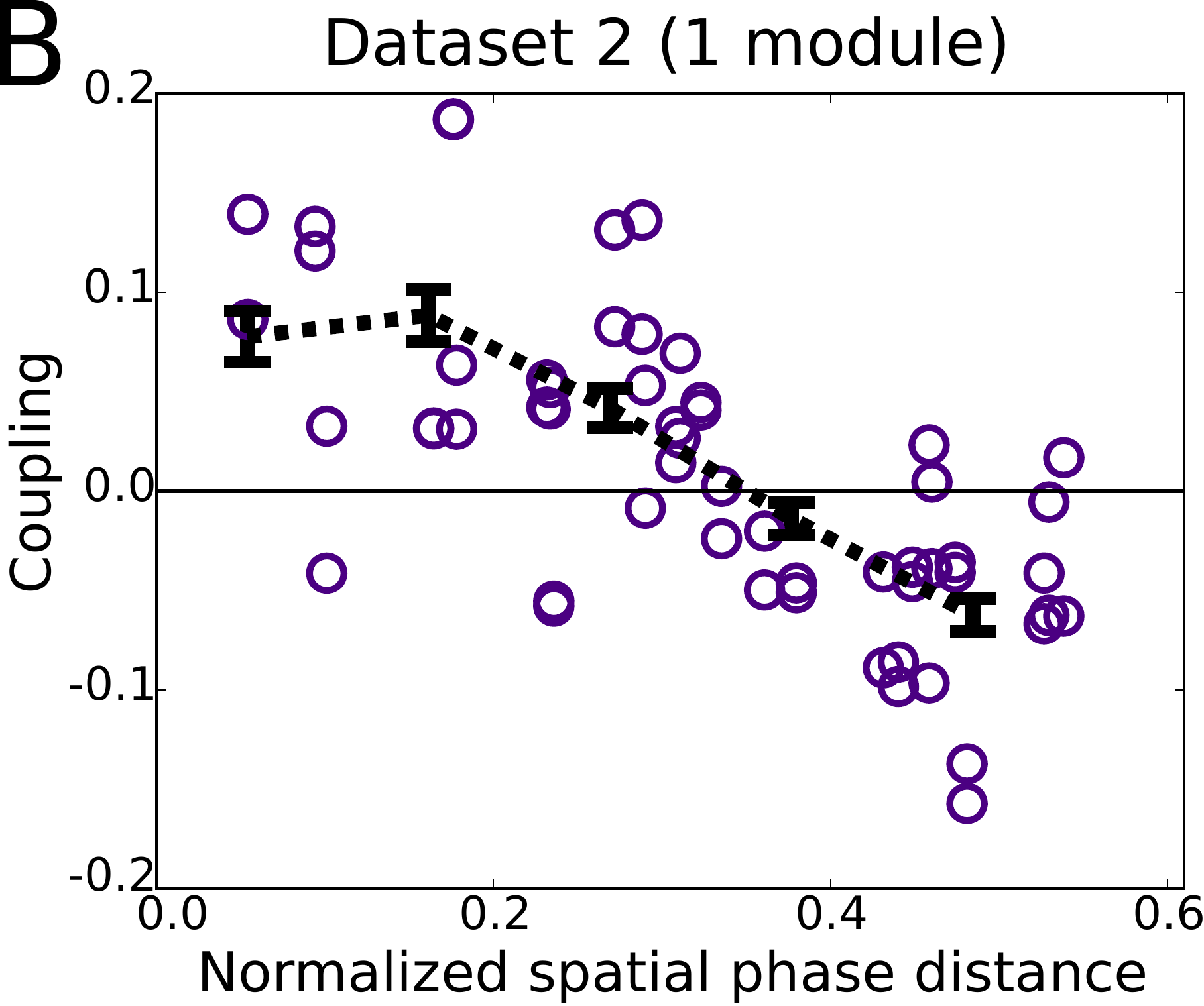}\\ \\
\includegraphics[width=48mm]{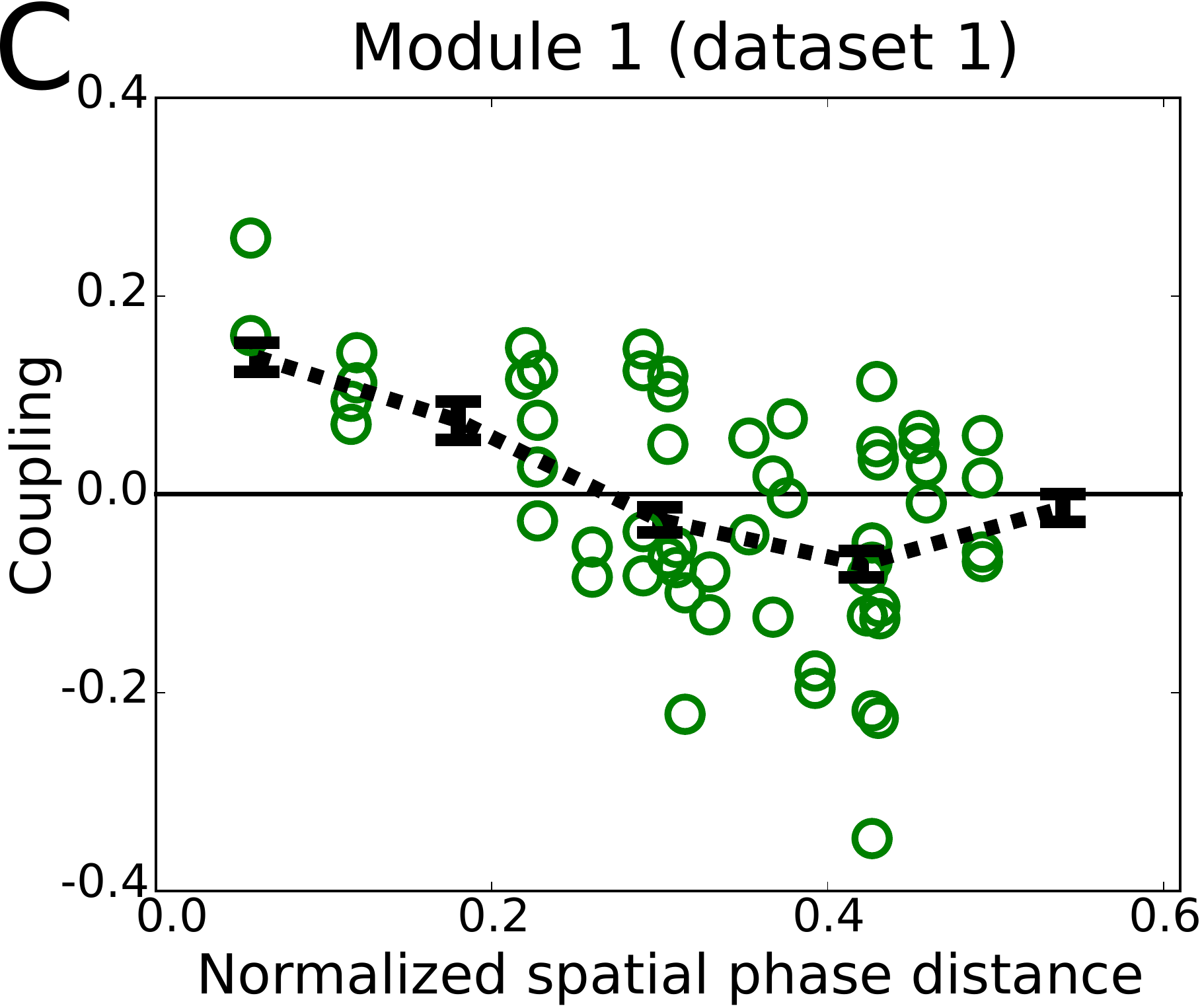}
\includegraphics[width=48mm]{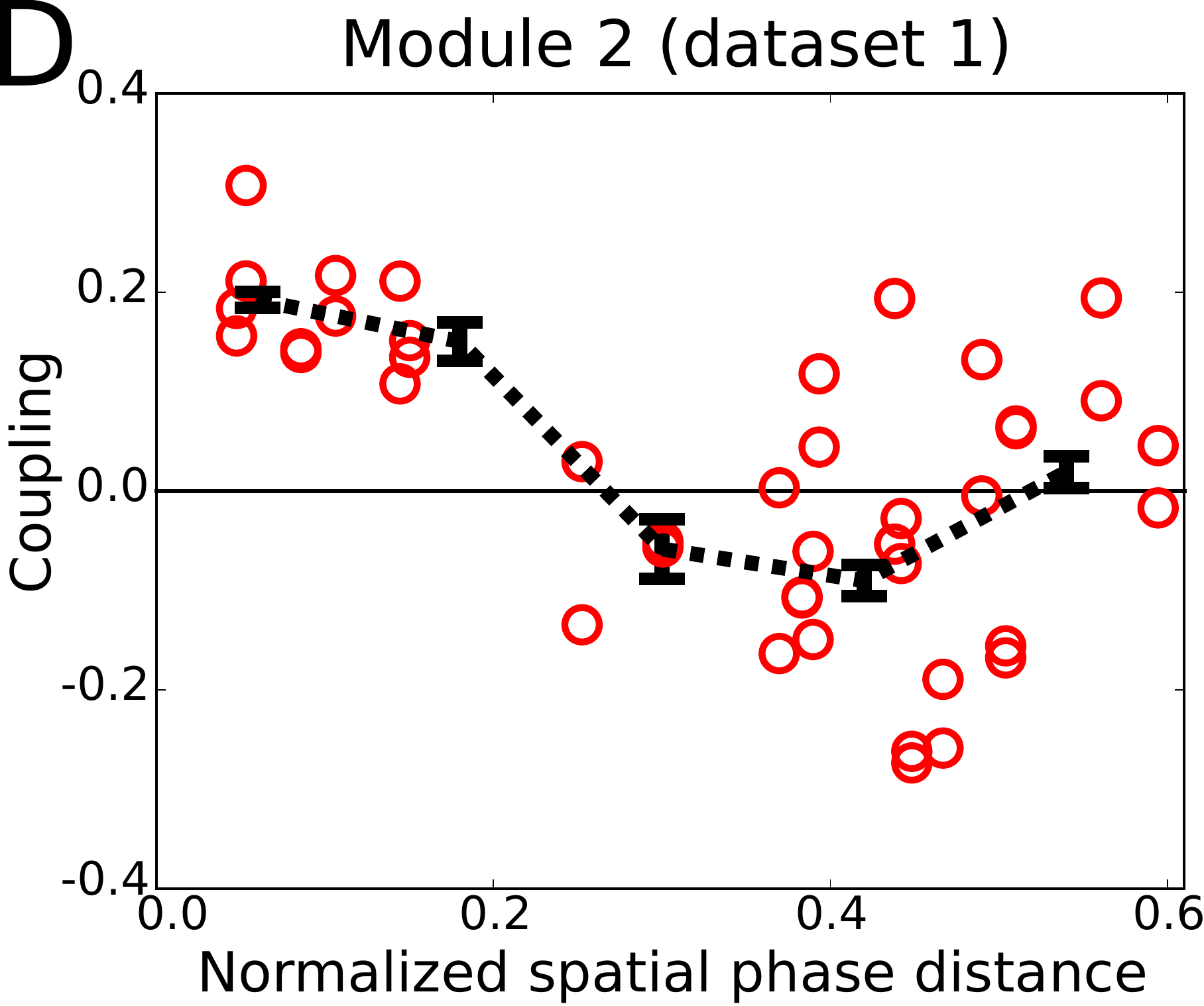}
\includegraphics[width=48mm]{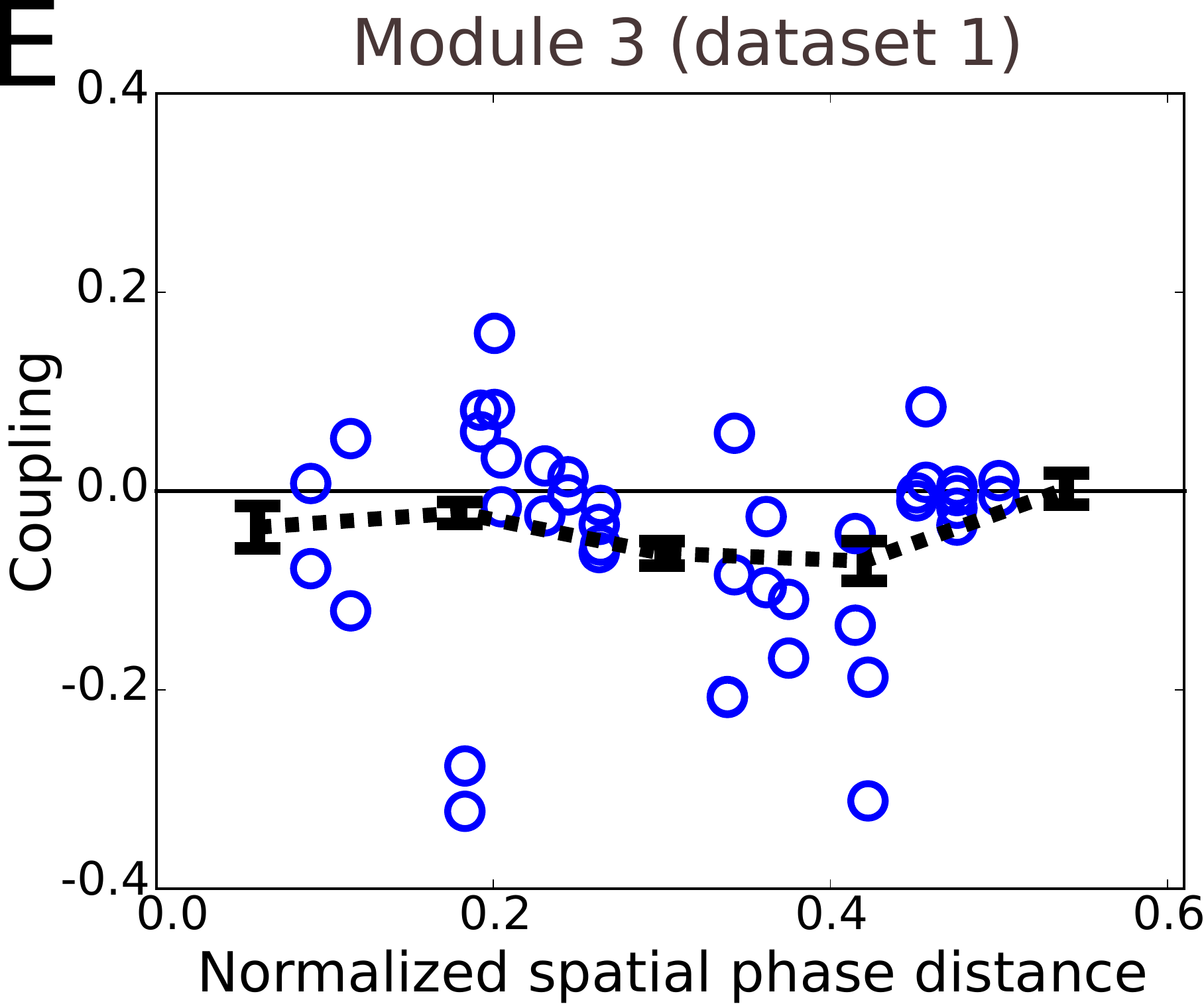}
  \end{tabular}
\end{center}
\caption{{\bf Couplings versus phase distance.} \small{Inferred couplings are positive for small phase differences while they become negative 
for larger phase separations, both for data set 1 (A) and data set 2 (B). When we break the population to the three contributing modules 
of data set 1, this pattern persists for the smaller modules (C,D) while for the largest module (E) the excitatory part is absent.
In each plot, the circles represent the inferred values using the full data length. The black lines show the average 
values of the couplings calculated from 20 random partitions of the data.}}
\label{FigJsvsPhaseDist}
\end{figure}

Interestingly, no matter which of the various external fields we used, when neurons $i$ and $j$ both 
belong to one of the two smaller modules of data set 1, or the one module of data set 2, the inferred couplings, $J_{ij}$, showed a consistent 
dependence on the spatial phase difference, with nearby phases showing positive $J_{ij}$ while those further away more negative values. 
This is shown in Fig.\ \ref{FigJsvsPhaseDist} for both data sets for the case of the Gaussian fields. The slopes and intercepts of linear regression 
lines were all significantly different from zero, both for the full data and the 20 sets of random halves (t-test, P$<$0.02) for all figures except 
for Fig.\ \ref{FigJsvsPhaseDist}E, where the slope and intercept of linear regression were not significantly different from 0 (t-test, P$>$0.7). 
We remind that with the Gaussian fields, the correlations between two cells 
due to overlapping fields are explained away.

\begin{figure}[h!]
\begin{center}
\begin{tabular}{cc}
\includegraphics[width=56mm]{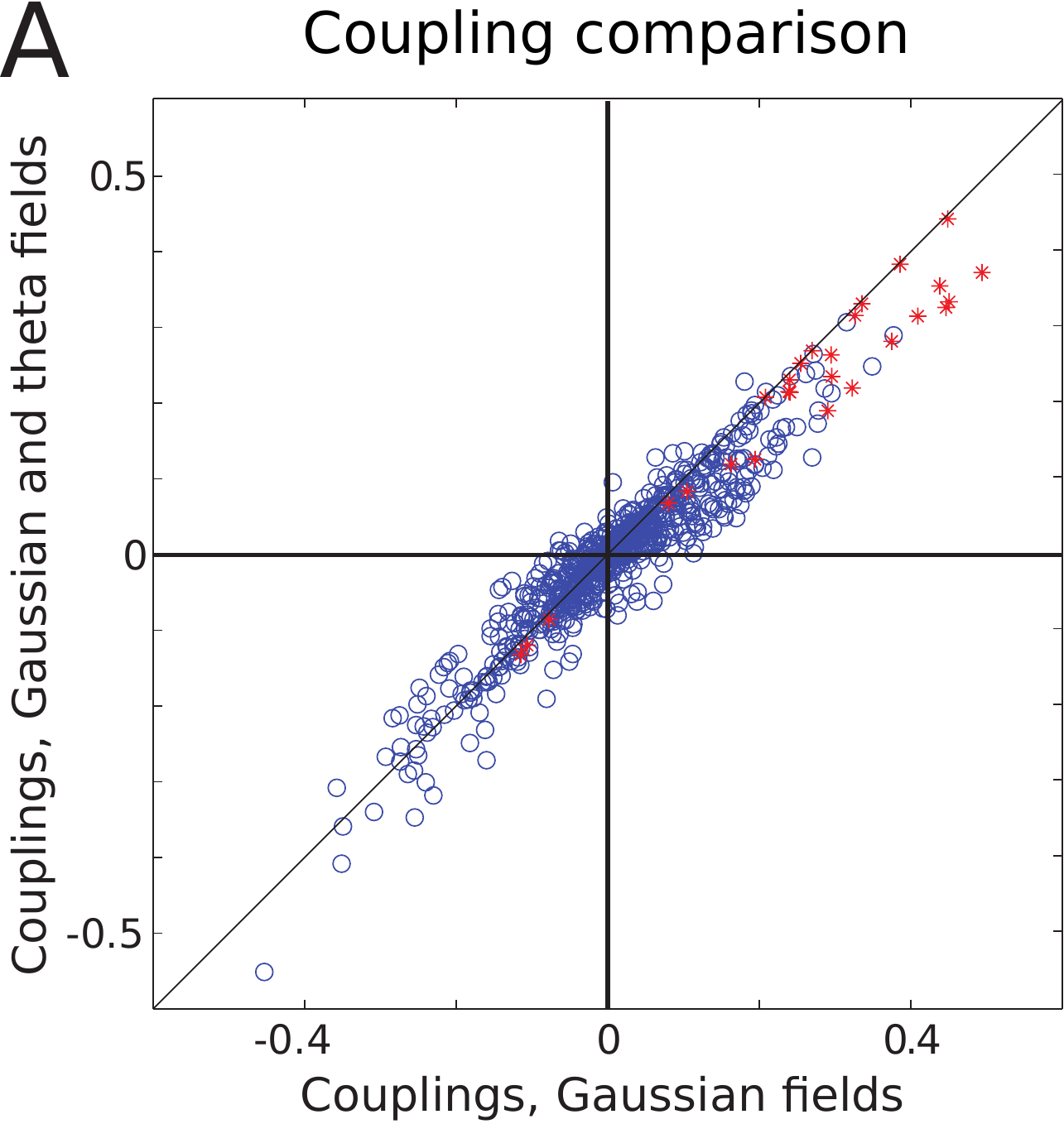} &
\includegraphics[width=57mm]{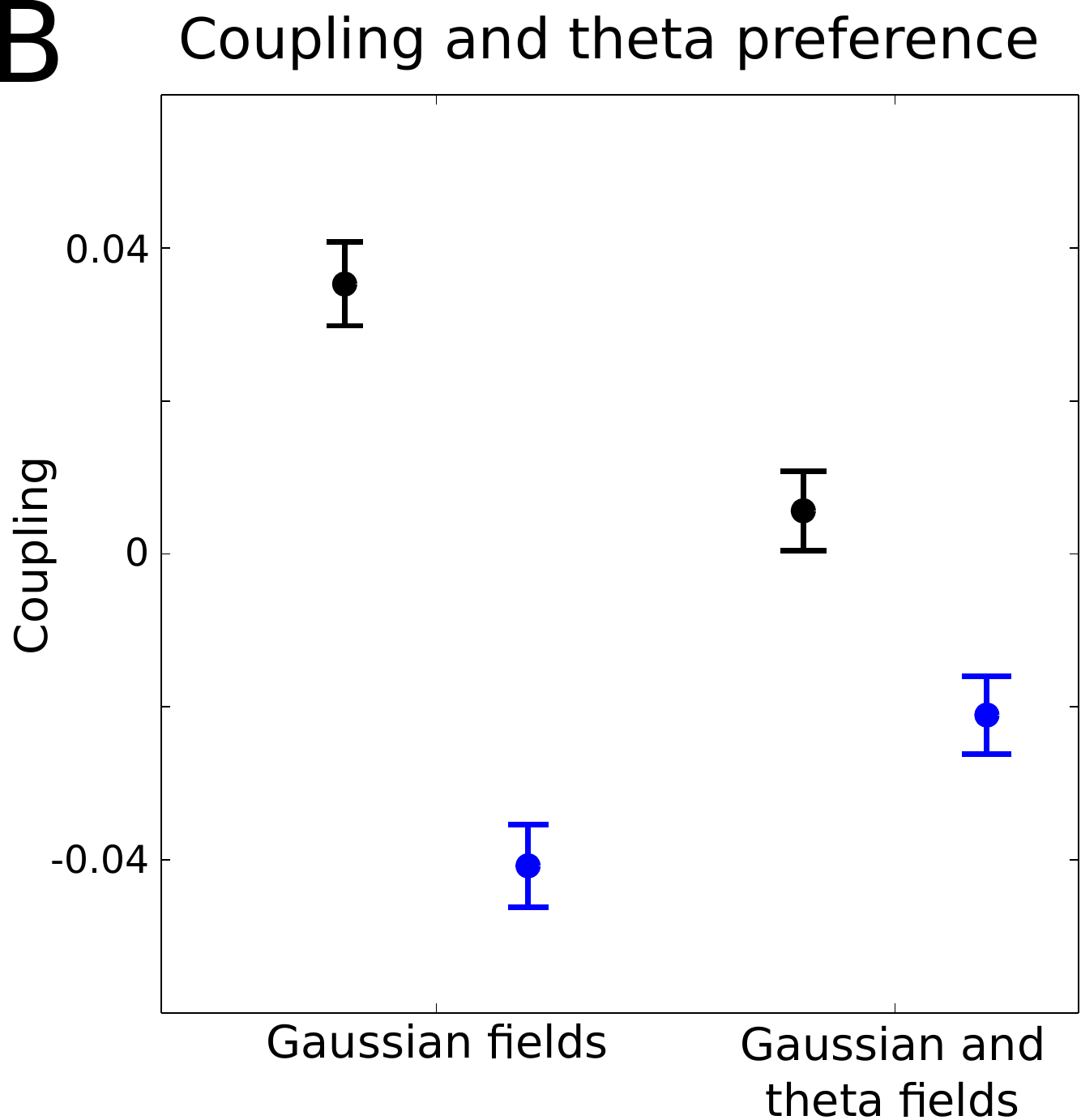}
  \end{tabular}
\end{center}
\caption{{\bf Effect of theta on the couplings.} 
  \small{(A) Adding theta to the Gaussian model has 
    little effect on the couplings (data set 1) with PCC, All = 0.95, PCC, SC = 0.97, PCC, NonSC = 0.94. 
(B) Mean of couplings from the two theta clusters in the Gaussian model with and without
theta included. Black: couplings between cells with similar theta phase preference. Blue: couplings between
cells with opposite theta phase preference. Error bars show the standard error of the mean. Without theta taken into account, the connections between 
cells that fire in the opposite theta phase are on average negative, while they are positive 
for those that tend to fire in the same theta phase. This difference is suppressed when theta is 
taken into account.}}
\label{ThHDAllvsconst}
\end{figure}

Since many cells in our data had some theta phase and head directional preferences, 
we also considered a model in which each cell was coupled to the head direction of the animal 
and the LFP theta oscillation through coupling constants that were inferred 
from the data; see Material and Methods. In general, there were only small differences between the couplings 
when theta and head direction were added. This can be seen 
in Fig.\ \ref{ThHDAllvsconst}A, which shows the couplings in the model with Gaussian fields 
with and without theta included. In this case, we observed a small but selective change, depending 
on the phase preference of the neurons. The cells could be clustered into two groups according to their theta phase preference (see Material and Methods): 
one with connections between cells of similar theta 
phase preference, and the other with connections between cells with opposite preference.  

Couplings between cells with similar theta phase preference were on average positive (average ($\mu$) significantly different from 0 (t-test, P$<$0.001)), 
whereas couplings between cells of opposite theta preference were on average negative ($\mu<0$, P$<$0.001).
As shown in Fig.\ \ref{ThHDAllvsconst}B, including the time-varying phase of theta as an external covariate resulted in 
shifting the coupling strength towards less positive values for pairs of cells that prefer the same phase of theta 
($\mu_{\text{no theta}} > \mu_{\text{theta}}$, P$<$0.001), whereas the opposite was true for 
couplings between cells that showed preference to opposite phase of theta ($\mu_{\text{no theta}} < \mu_{\text{theta}}$, P$<$0.001).
\begin{figure}[h]
\begin{center}
\begin{tabular}{cc}
  \includegraphics[width=58mm]{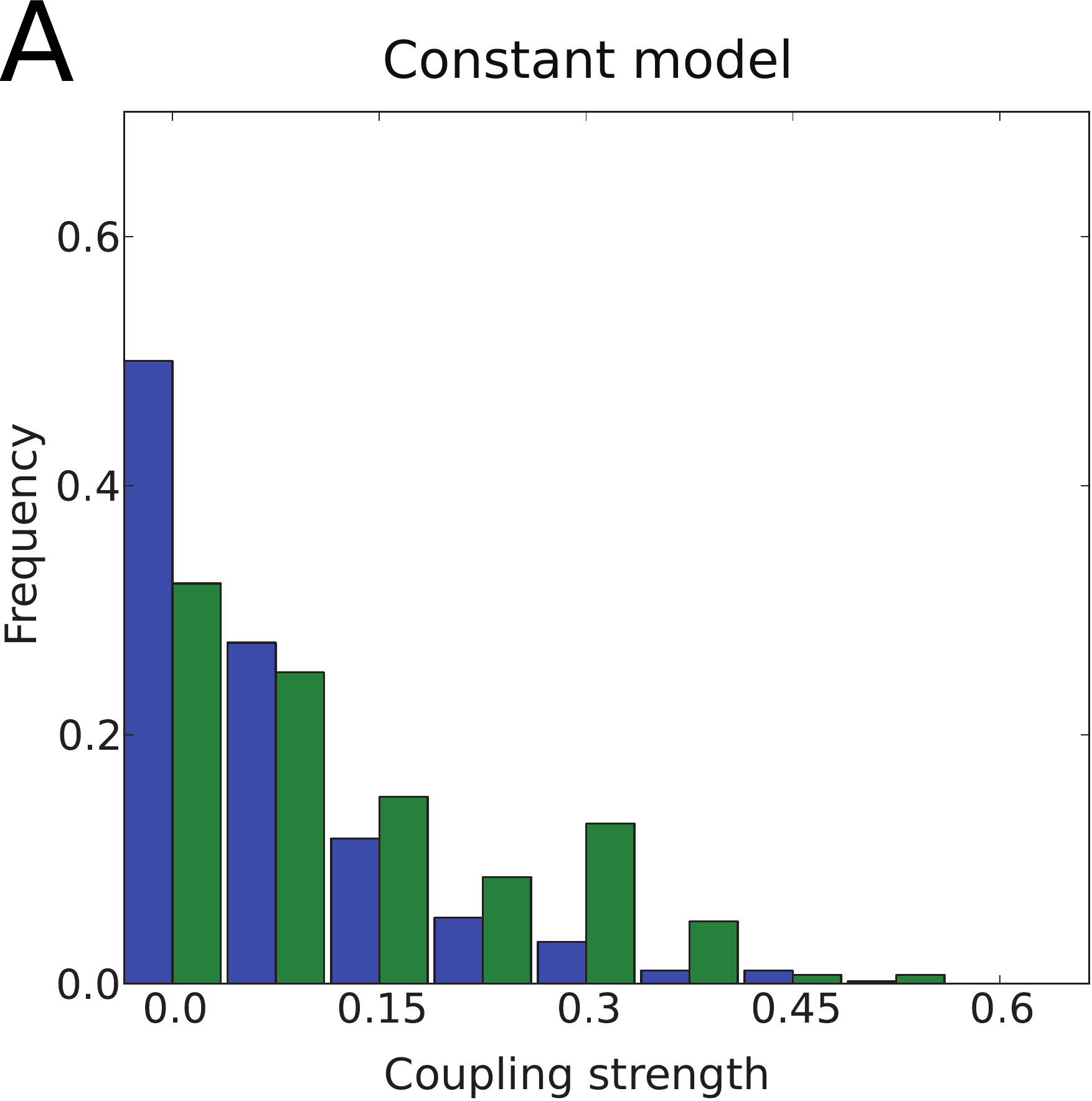} &
  \includegraphics[width=58mm]{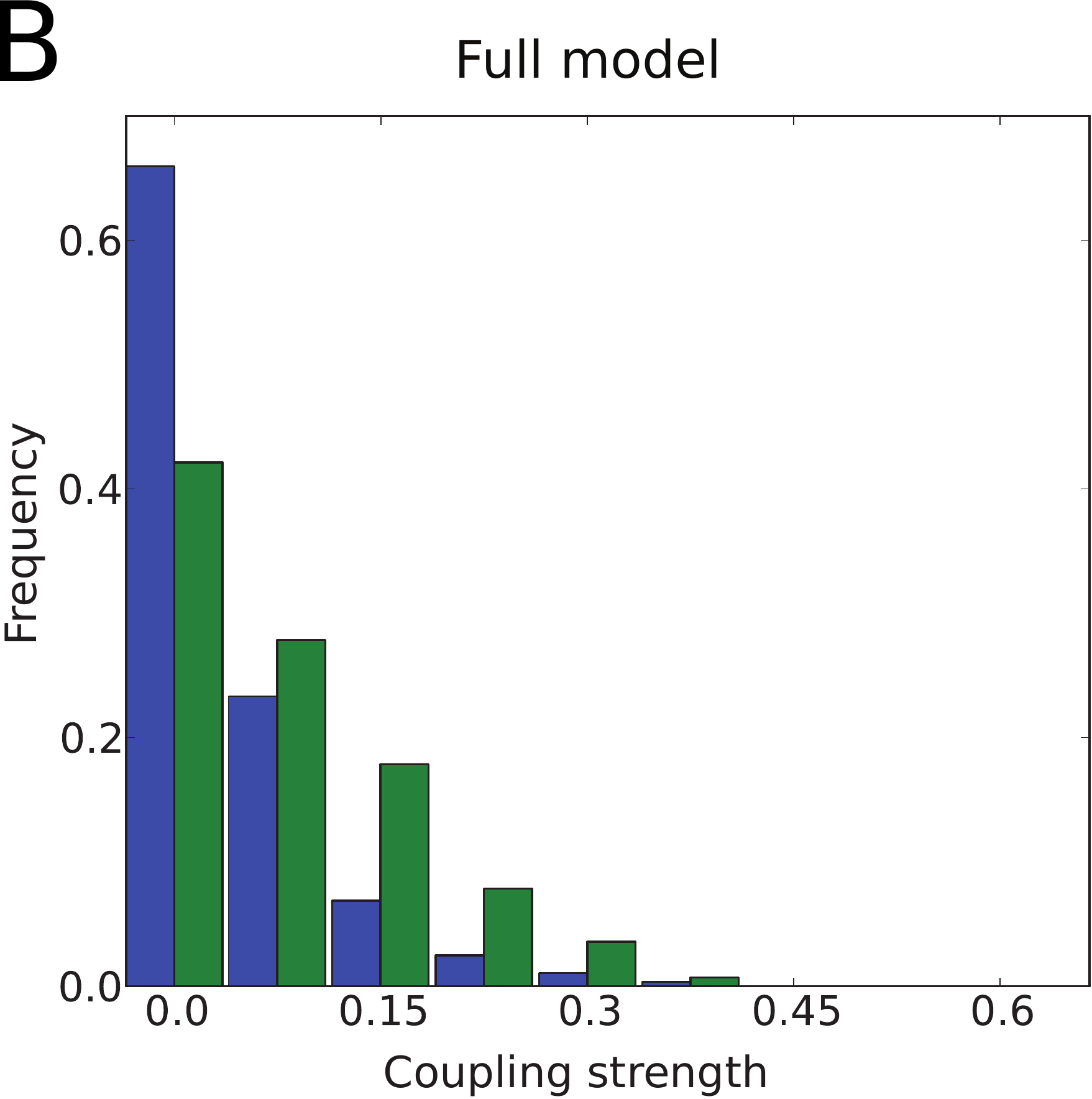}
\end{tabular}
\end{center}
\caption{{\bf Couplings between and within modules.} 
\small{Both couplings between and within modules have a mean value very close to zero. 
The probability of the absolute value of the couplings for the model with constant (A) and 
full (B) fields are shown here.
For the between module couplings (blue bars) there is a bigger peak at zero compared to the within module couplings (green bars), and the 
green histogram has bigger mass at larger values.}}
\label{FigWithinModule}
\end{figure}

One would expect, based on the experimental indications of modules operating independently, 
that grid cells of the same module are more likely to participate in the same functional network 
than neurons from different modules. 
We found that the couplings within and between modules in data set 1 both had means close to zero (within modules (mean$\pm$std): $-$0.01$\pm$0.13, 
between modules: $-$0.01$\pm$0.09). However, the within module couplings had a greater variance ($S^2_{\text{within}} > S^2_{\text{between}}$, 
P$<$0.001)), i.e. there was a higher proportion of couplings with high absolute values within modules than between, as can be seen in Fig.\ 
\ref{FigWithinModule}. This result was found to be stable with respect to data limitations, as shown in the next section. 

\subsection*{Stability of the couplings}
In this section we consider a number of factors that could have influenced our estimations of the couplings, 
and show that our results were stable with
respect to these factors.
\begin{figure}[h!]
\begin{center}
	\includegraphics[width=50mm]{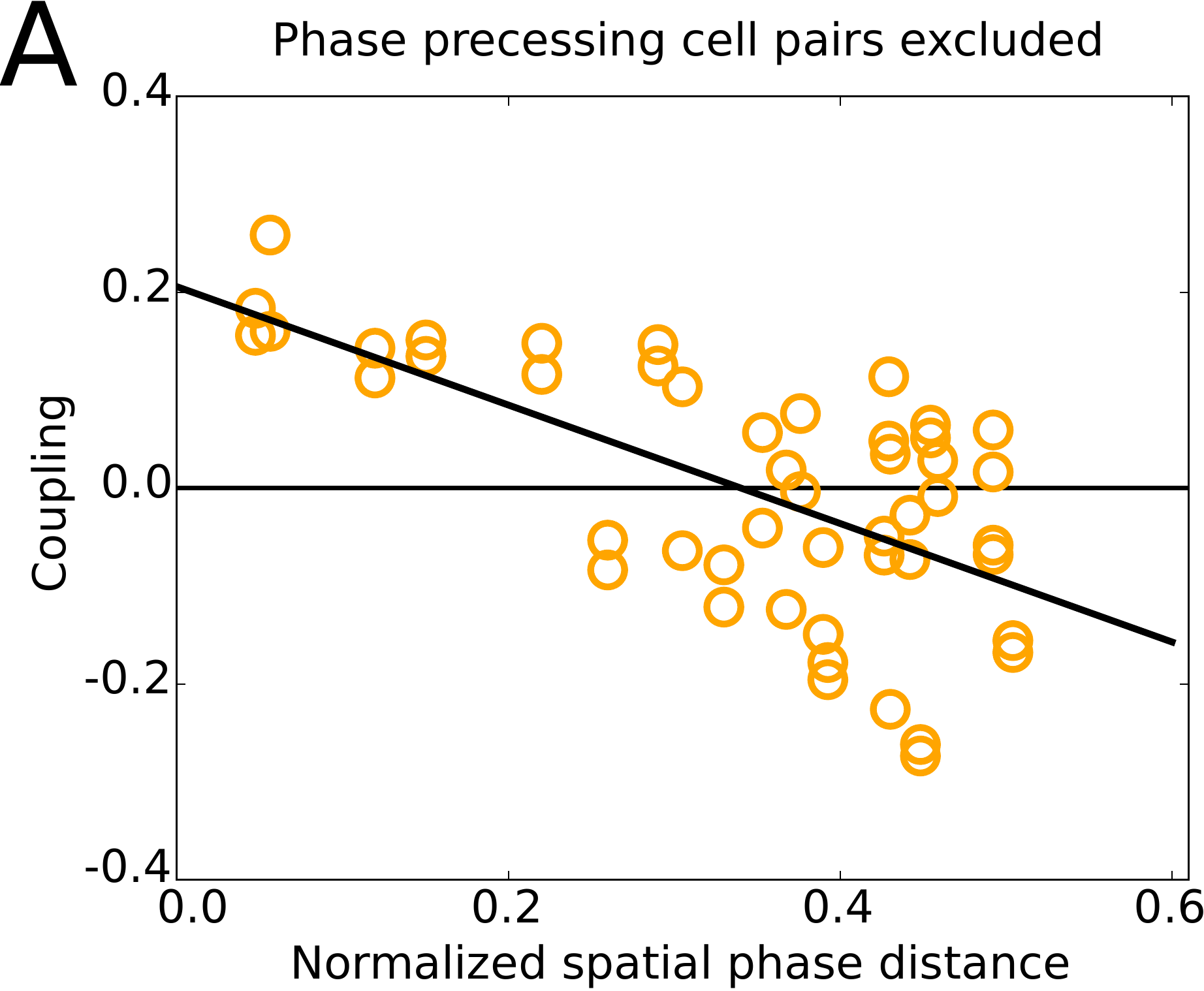}
	\includegraphics[width=50mm]{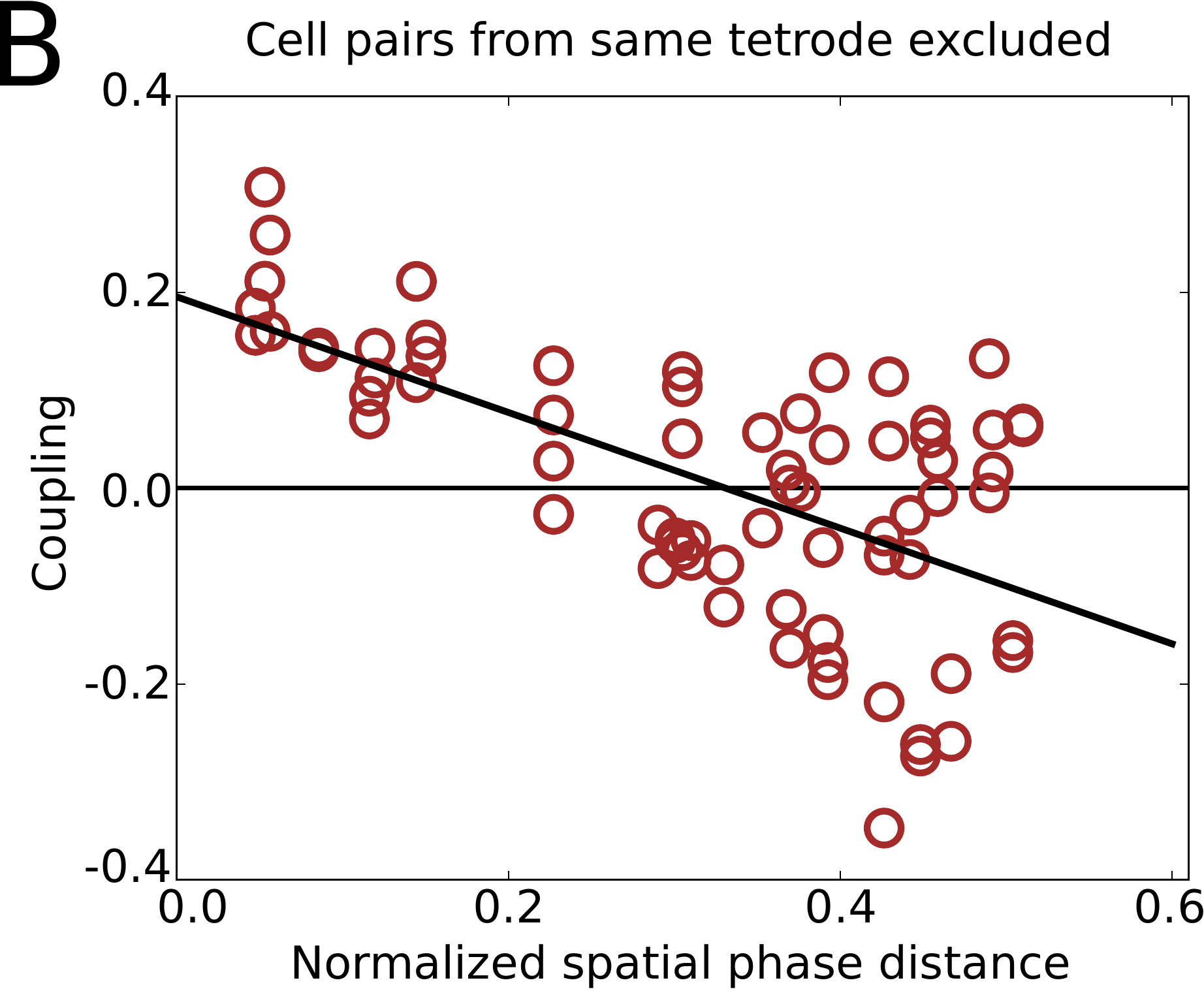}\\ \vspace{0.5cm}
	  \includegraphics[width=50mm]{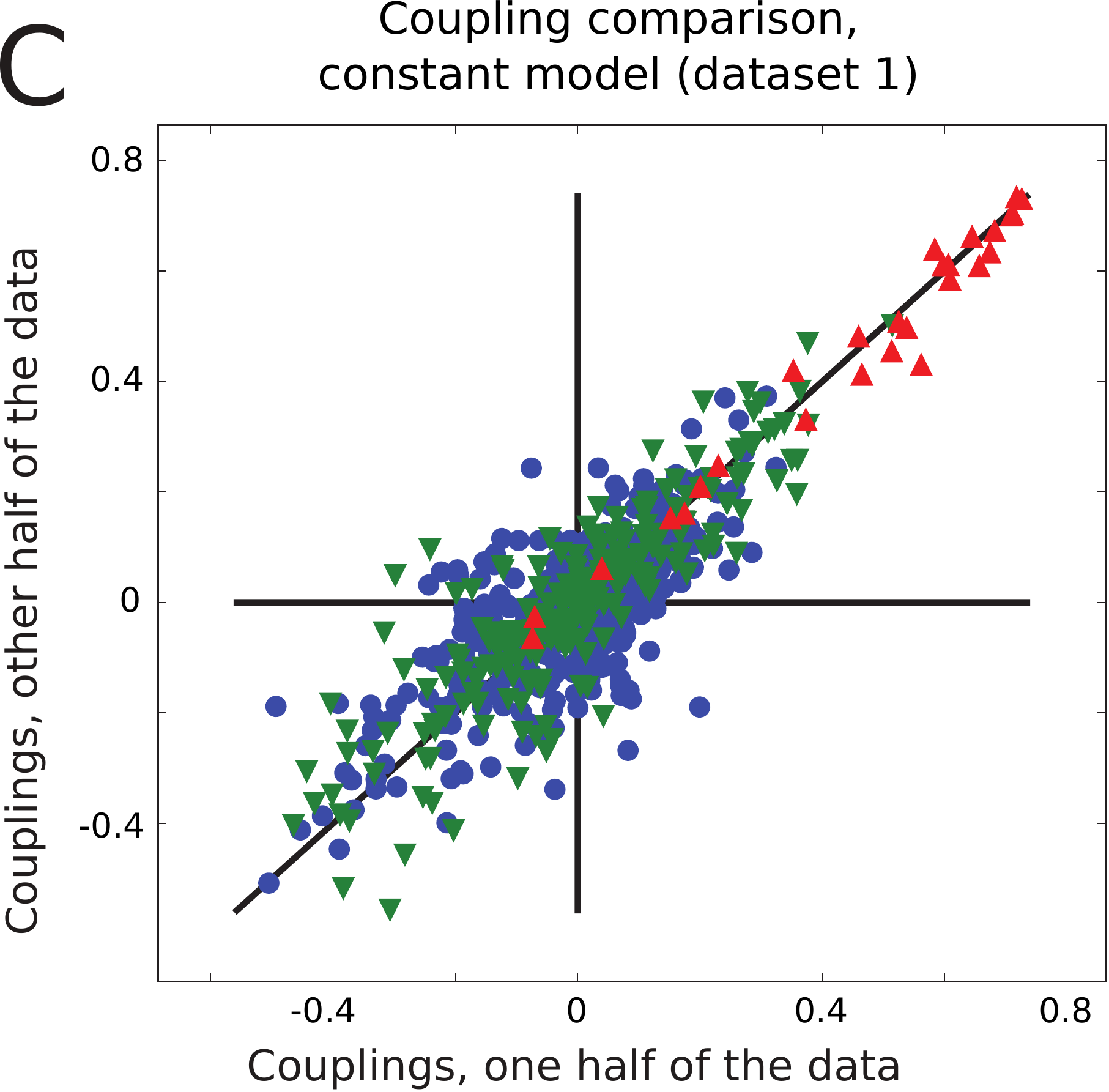} \:
  \includegraphics[width=50mm]{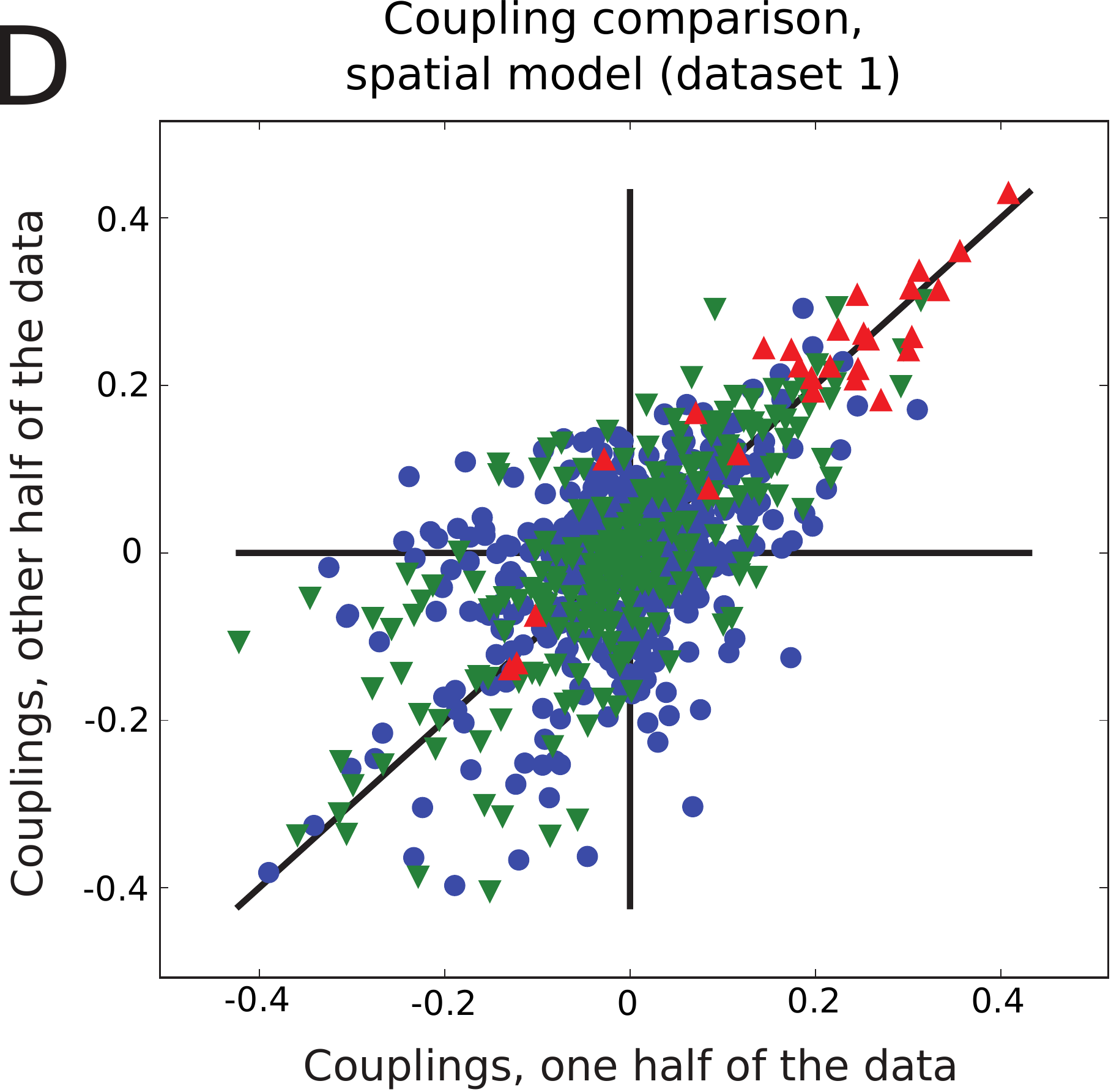}
    \includegraphics[width=47.5mm]{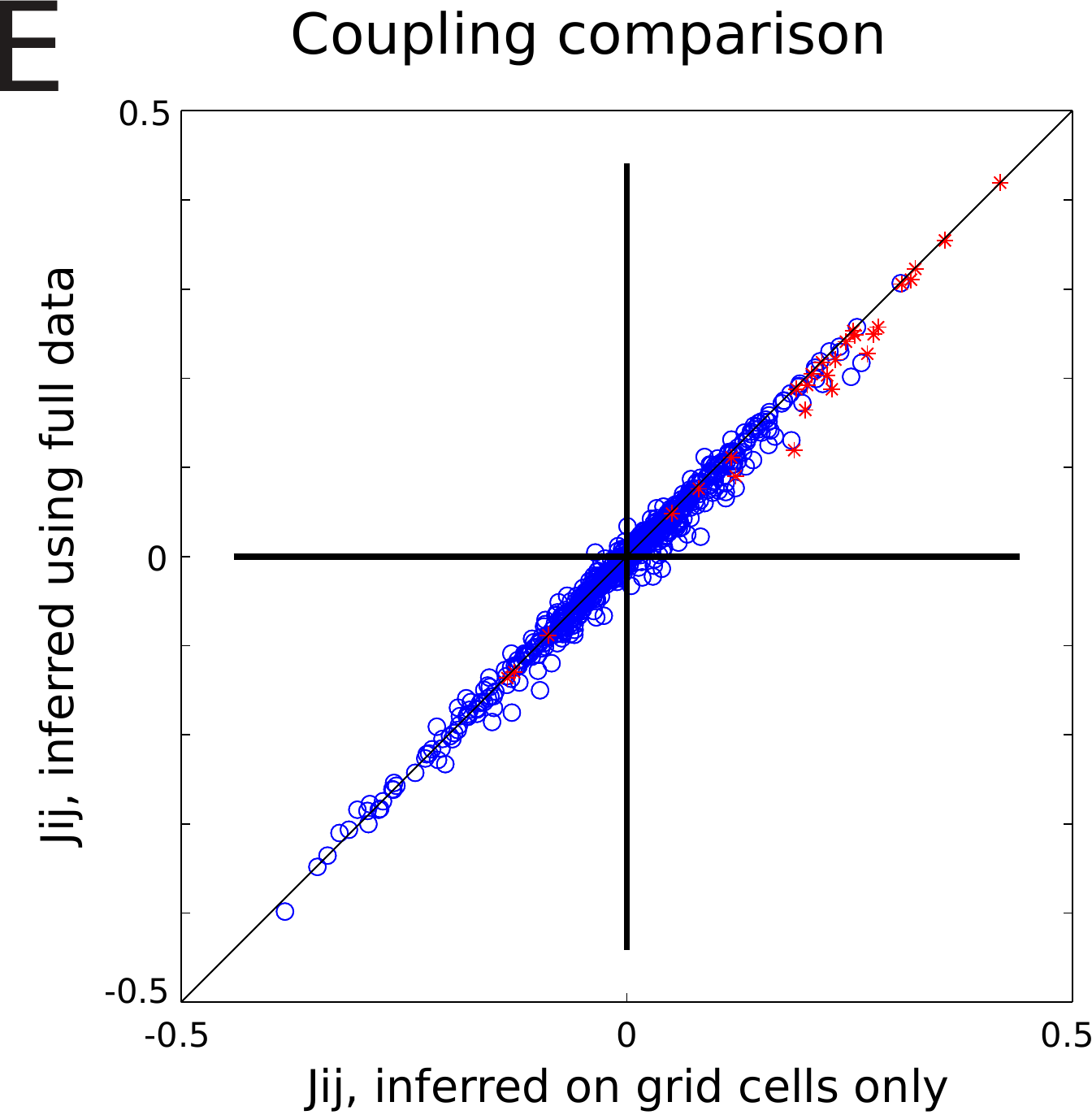}
\end{center}
\caption{\textbf{Stability of the inferred couplings.} 
\small{Stability of the phase-dependent trend in inferred couplings filtered for 
cell pairs where at least one cell is phase precessing (A), as well as for couplings filtered for cells on the same tetrode (B). The phase dependence 
of the coupling can be seen to be similar to when all pairs were included. Couplings inferred using one random half of the data plotted against those inferred from the other half, assuming constant external field (C) or Gaussian spatial fields (D). The within module couplings (green triangles) 
consistently show more stability across partitions of the data
than the between module couplings (blue circles), but not as much as the self-couplings (red triangles). 
A: PCC, within modules = 0.88, PCC, between modules = 0.73, 
PCC, SC = 0.99. B: PCC, within modules = 0.73, PPC, between modules = 0.51, PCC, SC = 0.94. (E) The effect on between grid cells-couplings from including non-grid cells in the inference for the biggest data set (data set 1, 65 cells) is small.}
\label{fig:Jvsphase_precession}}
\end{figure}

It is known that some grid cells show phase precession. This could be an additional source 
of correlation, so we tried to address how phase precession can influence the couplings. We first investigated whether or not any of the cells in 
our data phase precess, focusing on data set 1. In general, quantifying phase precession in two-dimensions is a difficult task due to 
the changes in the animals movement direction within the field. To classify cells 
as phase precessing or not, we thus used a novel approach described in \cite{jeewajee2014theta}, correlating the distance to the field peak 
projected onto the current running direction with the phase of theta at the time of spikes. Our analysis revealed that 
13 of the 27 grid cells showed significant phase precession (5 of 8 in module 1, 6 of 7 in module 2, and 2 of 7 in module 3).
We then excluded the couplings between phase 
precessing cells from the analysis for the two smaller modules and found that this did not remove the trend 
reported in Fig. \ref{FigJsvsPhaseDist} between the spatial phase difference and the inferred couplings. As can be seen in Fig.\ \ref{fig:Jvsphase_precession}A, 
there was still a significant negative relationship  between coupling value and spatial phase distance for cell pairs in which at least 
one of the cells do not show significant phase 
precession (both the slope ($\hat{\beta}$ = $-$0.60) and intercept ($\hat{\alpha}$ = 0.21) of the linear regression line are significantly different from 0 (t-test, P$<$0.001)).
\pagebreak

It has been suggested that correlations and thus inferred couplings from multi-electrode recordings can be biased due to problems with spike sorting 
\cite{ecker2010decorrelated, ventura2012accurately}. Since the main part of our conclusion is on the phase dependence of the correlations and functional 
connections and not their actual value, and since the phase of grid cells appears to be not anatomically ordered, it is unlikely that a phase dependent 
bias would be introduced to the correlations due to mistakenly assigning spikes to wrong cells. In addition to this, the cells in the two data sets 
analyzed here were recorded using hyperdrives that consist of 14 independently movable tetrodes \cite{Stensola2012}. It has been suggested that a tetrode 
is unlikely to record signals from cells farther than 65\textmu{}m away \cite{gray1995tetrodes}. As the distance between tetrodes on the hyperdrive is approximately 
250$\pm$50 \textmu{}m, it was very unlikely that the same cell was recorded on two tetrodes, and in 
that way confound our results across tetrodes. We therefore 
examined the couplings versus spatial phase for cell pairs from different tetrodes, and found that this led to a qualitatively similar result, as shown 
in Fig.\ \ref{fig:Jvsphase_precession}B (both the slope ($\hat{\beta}$ = $-$0.59) and intercept ($\hat{\alpha}$ = 0.20) of the linear regression were 
significantly different from 0 (t-test, P$<$0.001)).

In order to investigate the stability of the inferred couplings and the various covariates 
to data limitations we inferred the parameters of the models using only half of the data, 
and compared them with the ones from the other half. For this, we defined the spike data as being made up of consecutive time pairs, $\big(\mathbf{S}(t),\mathbf{S}(t+1)\big)$
and created partitions by randomly selecting
$50\%$ of the pairs. In this way, we generated 20 random sets, and for each set inferred the couplings using constant fields without 
taking theta and head direction into account, and Gaussian fields with theta and
head directional input included (the full model).
In general, the inferred couplings from these random halves were correlated with each other. 
As shown in Fig.\ \ref{fig:Jvsphase_precession}C and D, the within module couplings were more stable than the between-module ones, 
with an average Pearson correlation coefficient of $0.88$ versus $0.73$
for the constant field model, and $0.70$ versus $0.51$ for the full model.
We noticed that the self-couplings are the ones 
that are most stable from one half to the other, showing a Pearson correlation coefficient 
of $0.94$ between the couplings inferred from the two halves for the full model. 
We also found that the mean absolute values of the within and between module couplings maintained their relationship, 
with stronger couplings between cells within module than those between modules, for all 20 random partitions of the data 
($S^2_{\text{within}}>S^2_{\text{between}}$, P$<$0.005 for all 20 random partitions, in both constant field model and Gaussian field model).

The analyses reported here were produced using the data from two recordings of grid cells, the biggest of them consisting of 27 grid cells. 
This was the biggest data set we had access to, but still represents 
only a small fraction of the true local cell population.
One might wonder how much the connections between these cells would be influenced if we 
had access to recordings from more cells. As described in 
Material and Methods, data set 1 included neurons which were not classified as grid cells. 
We found that using this entire data set (65 cells) did not affect the couplings between grid cells 
(see Fig.\ \ref{fig:Jvsphase_precession}E).

\begin{figure}[h!]
\begin{center}
\begin{tabular}{cc}
  \includegraphics[width=50mm]{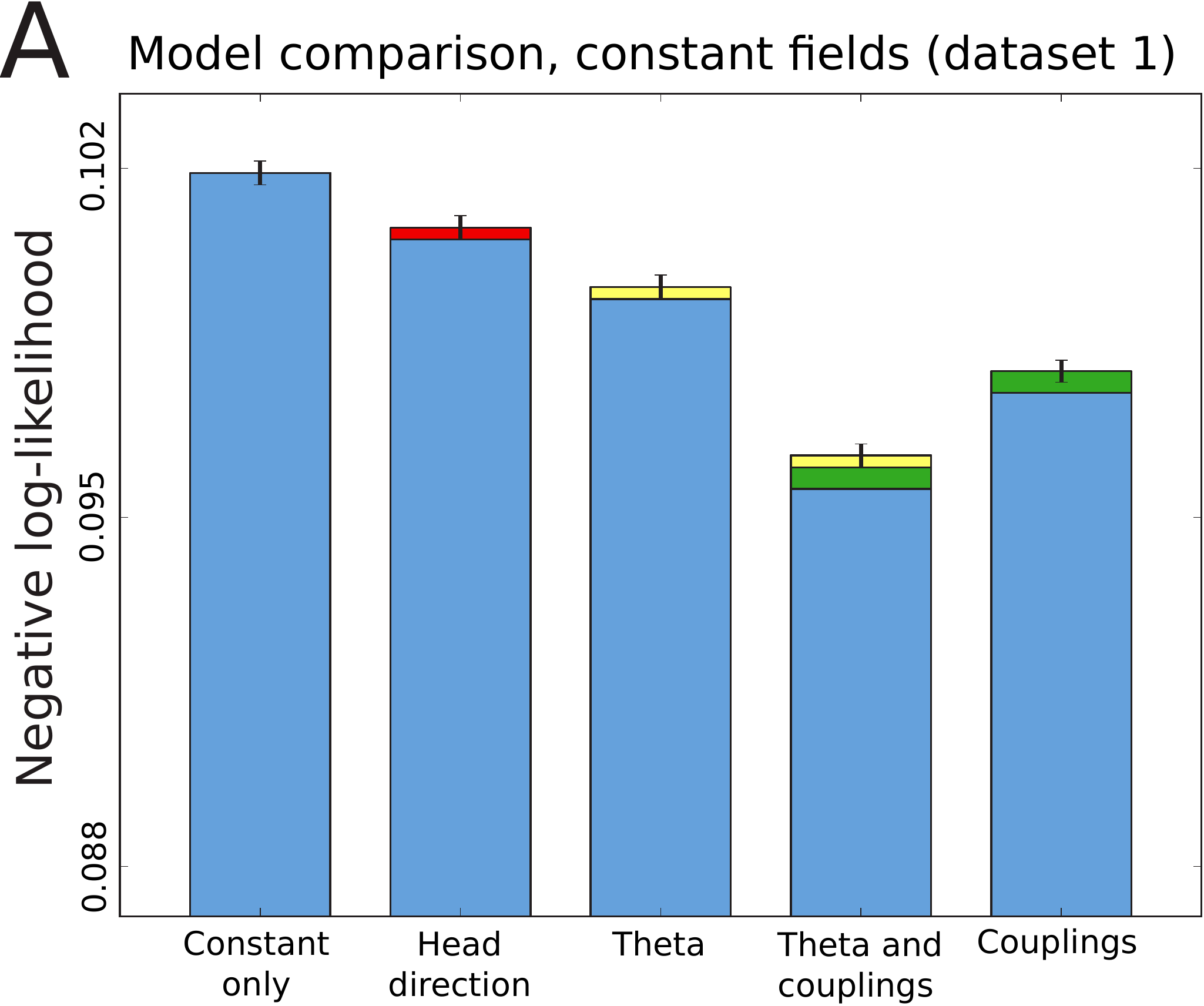} &
  \includegraphics[width=50mm]{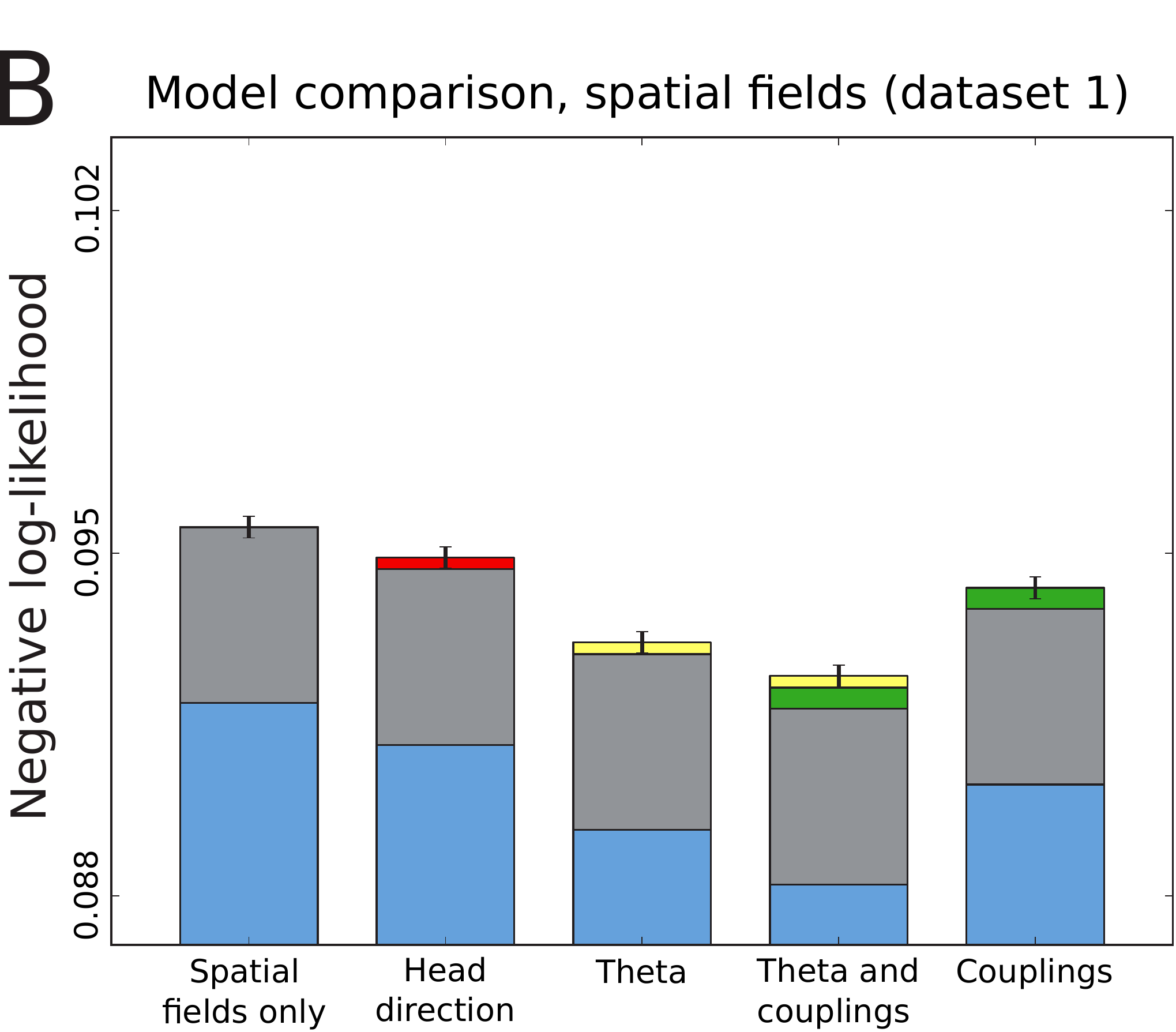} \\
  \includegraphics[width=50mm]{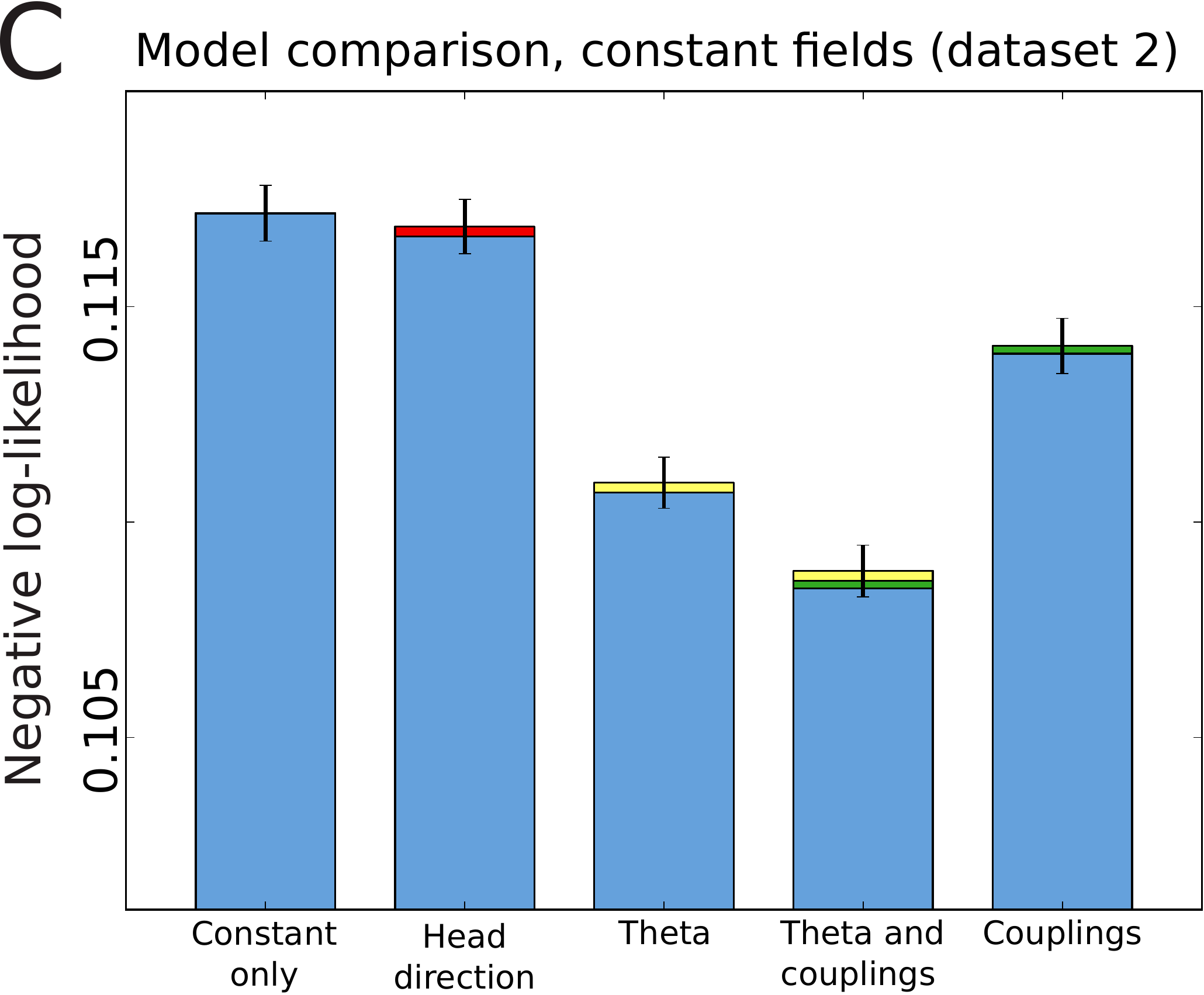} &
  \includegraphics[width=50mm]{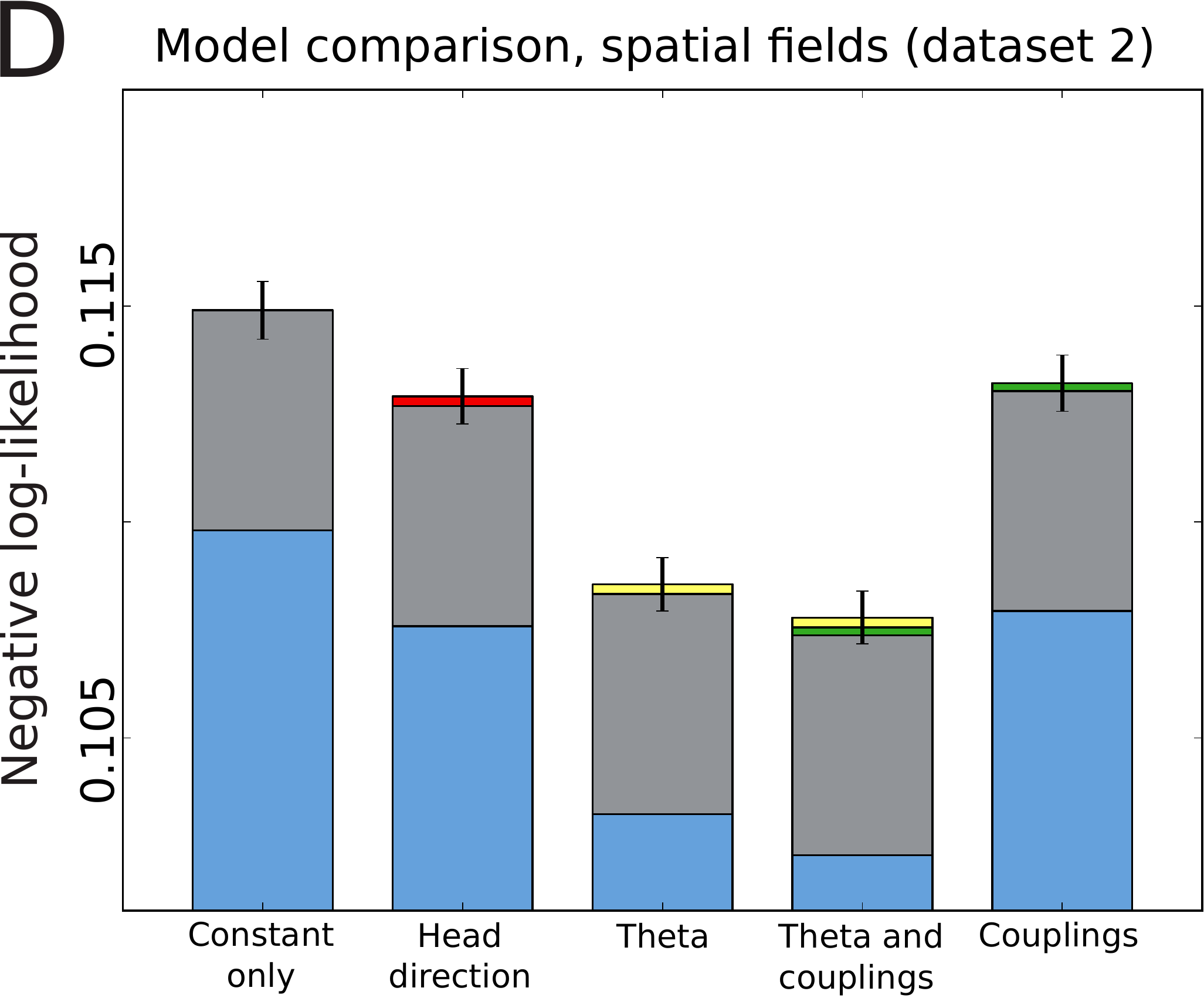} 
\end{tabular}
\end{center}
\caption{{\bf Statistical importance of the parameters.}  
\small{The negative log-likelihood per cell per time bin for the constant field model (A: data set 1, C: data set 2) and Gaussian field model (B: data set 1, D: data set 2) with different covariates included. 
Smaller values correspond to better explanatory power. The blue segment of the bar shows the negative log-likelihood. Adding parameters 
to a model will yield a log-likelihood-value greater than or equal to the model with fewer parameters. To avoid overfitting 
by including parameters, we performed an Akaike correction on the log-likelihood (see Material and Methods). The value of the Akaike-correction is 
shown for each covariate on top of the negative log-likelihood (blue) for each model: head direction (red), 
theta preference (yellow), 
and couplings (green). In (C, D), grey is the Akaike-correction due to the Gaussian spatial fields. These two plots show that adding the 
couplings always increases the explanatory power of the model, e.g. for the model with theta including couplings reduces the negative 
log-likelihood more than the penalty from the Akaike-correction for the added number of parameters.}
\label{FigConsistJhalves}}
\end{figure}

\subsection*{Statistical importance of the couplings and covariates}
In order to evaluate the strength of the statistical effect of the couplings and the external covariates on explaining the correlations 
in spike trains, we calculated the log-likelihood of half of the data using parameters inferred from the 
other half for various models for both data sets. The results are shown in Fig.\ \ref{FigConsistJhalves}A-D. 
To correct for the number of parameters, the total log-likelihood was penalized according to the Akaike correction, 
that is by subtracting the number of inferred parameters (covariates and couplings) used in each model (see Material and Methods) \cite{Akaike}.
The negative log-likelihoods of the models without the couplings are also shown. In a likelihood ratio test, all covariates gave a significant increase 
($P<0.001$) compared to the constant field model. This was also the case where we included the couplings 
in each of the models compared to the same model without couplings.
In general, adding head direction as a covariate had little effect on the likelihood. The effect was even weaker when including speed as
a covariate, or using running direction instead of head direction (see methods), 
with the penalty from the Akaike correction larger than the increase in likelihood from the inclusion of the parameters. 
For the case of constant fields, adding couplings and then theta had the 
most significant effect.  It is interesting to note that, when comparing the constant field model to the 
model with spatial fields, the impact on the likelihood from including the couplings is reduced, as would 
be expected by explaining away the spatial component of the correlations. Adding theta resulted 
in a consistent increase in the log-likelihood yielding 0.0025 for the model with constant fields 
and 0.0026 for spatial.

\section*{Summary and Discussion}

What is known about the connectivity in the grid cell network is primarily based on anatomical {\it in vitro} studies. 
Recent studies show that stellate cells in layer II are connected to each other primarily through inhibitory interactions \cite{Couey2013,pastoll2013feedback},
and that the inhibitory drive varies dorsoventrally as the size of the grid spacing changes \cite{beed2013inhibitory}. 
As opposed to the connections between layer II stellate cells, within-layer recurrent excitation has been found between the main 
type of principal cells, namely pyramidal cells, in both layer III and V \cite{Dhillon2000}.
Although the picture drawn by these studies emphasizes the role of recurrent interactions
in developing the properties of grid cells, it does not show how interactions between grid cells quantitatively depend on properties such as theta rhythmicity and spatial phase separation, properties that play a major role in computational models of grid cells. A previous work on {\it in vivo} recordings that studied phase dependence of the interactions between cells in MEC focused on pairwise correlation analysis by using recordings from one dimensional tracks \cite{mathis2013multiscale}, showing that cells with nearby phases have stronger correlations than those far apart in phase. Another recent {\it in vivo} study used strongly peaked cross-correlations as a signal for the presence of connections and has concluded that grid cells with a wide range of phases project to a given inhibitory neuron \cite{buetfering2014parvalbumin}. To analyse the multi-neuronal recordings in grid cells we took a different approach from previous studies: that of statistical inference. We used a kinetic Ising model and studied how functional connections depend on phase difference between grid cells, their level of theta modulation, speed modulation and head directionality, and the statistical role that these connections play in shaping multi-neuronal activity.

The kinetic Ising model that we used here for the inference is a model with minimal assumptions: (1) it is the maximum entropy distribution over the spikes of 
neurons at time $t$, given the spikes at time $t-1$ \cite{jaynes1982rationale}, and (2) it is pairwise (meaning it only takes into 
account the first-order non-trivial interactions). 
Being a generalized linear model, it is closely related to other GLMs used for analyzing population recordings from other parts of the 
brain \cite{Truccolo2005, pillow2008spatio,Rebesco2010}, and it also employs the maximum entropy approach used by many in analyzing 
neural \cite{shlens2006structure,schneidman2006weak} or other biological data \cite{lezon2006using}.  
Our analysis showed that the correlations and the functional connections between grid cells demonstrate a spatial phase dependence, even 
when spatial variations in rate (as well as other possible sources of correlations, such as theta oscillations and head direction) are taken into account.
Both correlations and functional connections were positive for small phase differences. Functional connections became negative, while the correlations approached zero, for larger spatial 
distances for cells in the one module in data set 2, and in the two smaller modules in data set 1. 
This connectivity provides support for a role played by attractor dynamics  as suggested by several modelling efforts 
\cite{McNaughton2006,Fuhs2006,Burak2009, Couey2013}. The trend in the phase dependence was, however, less clear in the third module in data set 1: the common inhibitory portion was represented, but we did not find any functional 
excitation between cells close in phase, possibly 
because of the lack of recorded cells with similar phase in this module. 
We also found that the absolute value of the couplings was bigger for pairs of cells that belonged to the 
same module than those  belonging to different modules. 
This supports the idea that neurons in the same module form a more coherent population of neurons, 
bound together in a stronger manner than those in different modules.

In attractor models of grid cells, the phase dependent connectivity pattern allows the network to maintain a continuum of stable states such that, 
if the neurons of a single module could be aligned according to their phases, the activity on that neural sheet would itself show a regular pattern 
of activity. This local and relatively rigid relationship between within-module grid cells has been surprisingly well supported. 
First identified in \cite{Hafting2005}, grid cells were found to locally share both orientation and spacing that were later observed to remap and 
deform coherently \cite{Fyhn2007, Stensola2012, yoon2013specific}. It has also been shown that the characteristics of the grid pattern of one cell
were more stable relative to other grid cells than with respect to local features of the environment \cite{yoon2013specific}. 
This was even more pronounced in novel environments where the individual fields were still changing significantly relative to the environment
while remaining relatively stable between cells \cite{yoon2013specific}, further suggesting 
that the coding of the grid cells is more coherent within the 
grid cell population than it is with the actual space it is encoding. 
Even more convincing, a recent study looking at a 
large population of cells taken from single animals in the same environment showed that the cells clustered into a finite number 
of modules \cite{Stensola2012} suggesting there exists not only the large number of cells necessary for an attractor map 
but that there might be a finite number of these networks working together to better provide a metric of space. 
Our work complements these studies in that we show that there exists the functional connectivity of the type necessary 
to establish the patterned network activity that has been proposed to explain the above experimental observations.

As opposed to the attractor model \cite{McNaughton2006,Fuhs2006,Burak2009}, other grid cell model frameworks, the oscillatory interference 
\cite{burgess2007oscillatory} and the adaption model \cite{kropff2008emergence}, were originally conceived as single cell models that suggest 
that the periodic firing comes from a combination of 
convergent input and cellular mechanisms within an individual neuron. As such, the role they have 
prescribed for the lateral connectivity has been mainly to align the grid patterns of the 
cells, without requiring
any phase dependence in the couplings per se. However, it has recently been noted \cite{si2012grid,moser2014grid} that in the adaptation model, interactions 
between grid cells can also be learned, resulting in a developmental model for the phase dependent connectivity which could later sustain a continuous attractor dynamics. 
In addition to aligning the grids, this connectivity will allow the adaptation model 
to code for novel environments much more rapidly while maintaining the stabilizing benefit of having convergent spatial input.

In our statistical inference, we considered various external covariates that comprise what 
is known about the single cell coding of these cells, including spatial, speed, theta oscillations, 
head direction and running direction inputs. Adding these additional covariates
to the models with constant field or Gaussian fields had little effect on the connectivity, but there was a significant
weakening of the couplings when we compared the couplings of the Gaussian model to those of the model with constant fields. 
This is not surprising, as a component of the correlations in the model with constant fields 
was likely due to overlapping fields which was better explained by the spatial component of the Gaussian model. 
One benefit of using a statistical model is the quantification of the relative contribution of the individual
covariates to the overall likelihood of the data under the model, with the spatial component having the strongest
impact followed by functional connectivity and theta preference.  
Speed, head direction and running direction, as covariates, had a small impact in all cases that
we considered. 

In all the statistical models, ranging from constant external field to Gaussian with and without 
theta and head direction, we found that the model without couplings was worse at 
explaining the statistics of the data than the same model with couplings, even when the 
Akaike corrections were taken into account. Further support for the significance of the 
couplings come from the stability of the connectivity when inferred from 
separate halves of the data.

Since the self-couplings appeared to be the most stable when one random partition of the data was 
compared to the other,  we wondered how the rest of the couplings would react if we 
did not include the self-couplings. With the refractory period in mind, positive 
self-couplings might seem counter-intuitive. However, the refractory period lasts for 
only a few milliseconds, and we use 10 ms time bins. In addition, grid cells are 
primarily active only when the animal is in the cell's spatial fields, and silent 
otherwise, i.e. the state of a grid cell in a time bin is likely to be equal to the 
state in the previous time bin, which a statistical model could interpret as a positive 
self-coupling. Removing self-couplings, however, had little effect on the couplings between cells (Pearson correlation coefficient $>$ 0.98 for the constant 
model and the full model, for both data sets).

Stellate cells of MEC layer II, the main grid cell candidates, are known to functionally inhibit each other. 
In our analysis, the inferred connections were both inhibitory and excitatory. There are a few points to note regarding this apparent contradiction. 
First, considering the recording locations of the tetrodes in data 
set 1 (see Supplementary figure 4 (rat 14147) in \cite{Stensola2012}), and that a 
number of cells in this data set show head direction preferences, a property rarely observed in the layer II population 
\cite{Sargolini2006}, many of these cells are most likely recorded from deeper layers where, as mentioned, 
both intra- and interlayer excitatory connections between principal cells have been found. For data set 2, on the other hand, it seems probable that a 
bigger fraction of the cells is from layer II (see Supplementary figure 14 (rat 13855) in \cite{Stensola2012}). It is, however, not possible
to confirm the exact location or principal cell type for the cells analyzed here. Second,
the relationship between the inferred functional connections and the underlying 
anatomical connectivity is a nontrivial one which may involve other non-recorded neurons. 
It is also possible that the correlations driving the functional connectivity come from a common input that was not accounted for
here. This input, however, should be non-spatial, non-directional and independent of theta phase, but still depend on the spatial 
phase difference between pairs of neurons and whether or not they belong to the same module.
It would be interesting to see what such
a signal could look like. The existence of such an input would, of course, leave the question open as to how the 
local network is connected, while opening a new possibility that the grid cell modules
play a role in encoding currently unidentified features that are neither spatial or directional. 
Since it is possible in computer simulations to identify the presence or absence 
of a synapse based on the inference of functional connections \cite{Hertz2011,capone2014inferring}, 
it would be very interesting to see how the inferred functional couplings and correlations 
look like for a data set exclusively from layer II cells for which the actual functional connectivity between stellate cells is known.
In addition, considering the fact that modules span layers \cite{Stensola2012}, our results also make a case for taking a closer 
look at the between layer connectivity and how the different cell types and connectivity patterns might work together to develop 
the grid cell code.

With Gaussian fields, the model with only theta has a slightly higher likelihood than the one with only couplings,
although the couplings still exhibit the phase dependence shown in Fig.\ \ref{FigJsvsPhaseDist}.
The relative improvement gained by pairwise connections in explaining the data is known to scale 
with the size of the recorded population \cite{schneidman2006weak, Roudi2009,stevenson2012functional}, while other 
sources of higher order correlations will also scale up. 
It would therefore be interesting to see how the relative contribution of the various 
factors, in particular that of theta oscillations, will scale compared to that of the pairwise couplings. 
Future large-scale recordings of grid cells should allow us to perform such analyses. 

\section*{Material and Methods}

\subsection*{Data}
\begin{figure}[h!]
  \begin{center}
	\begin{tabular}{cc}
      \includegraphics[width=50mm]{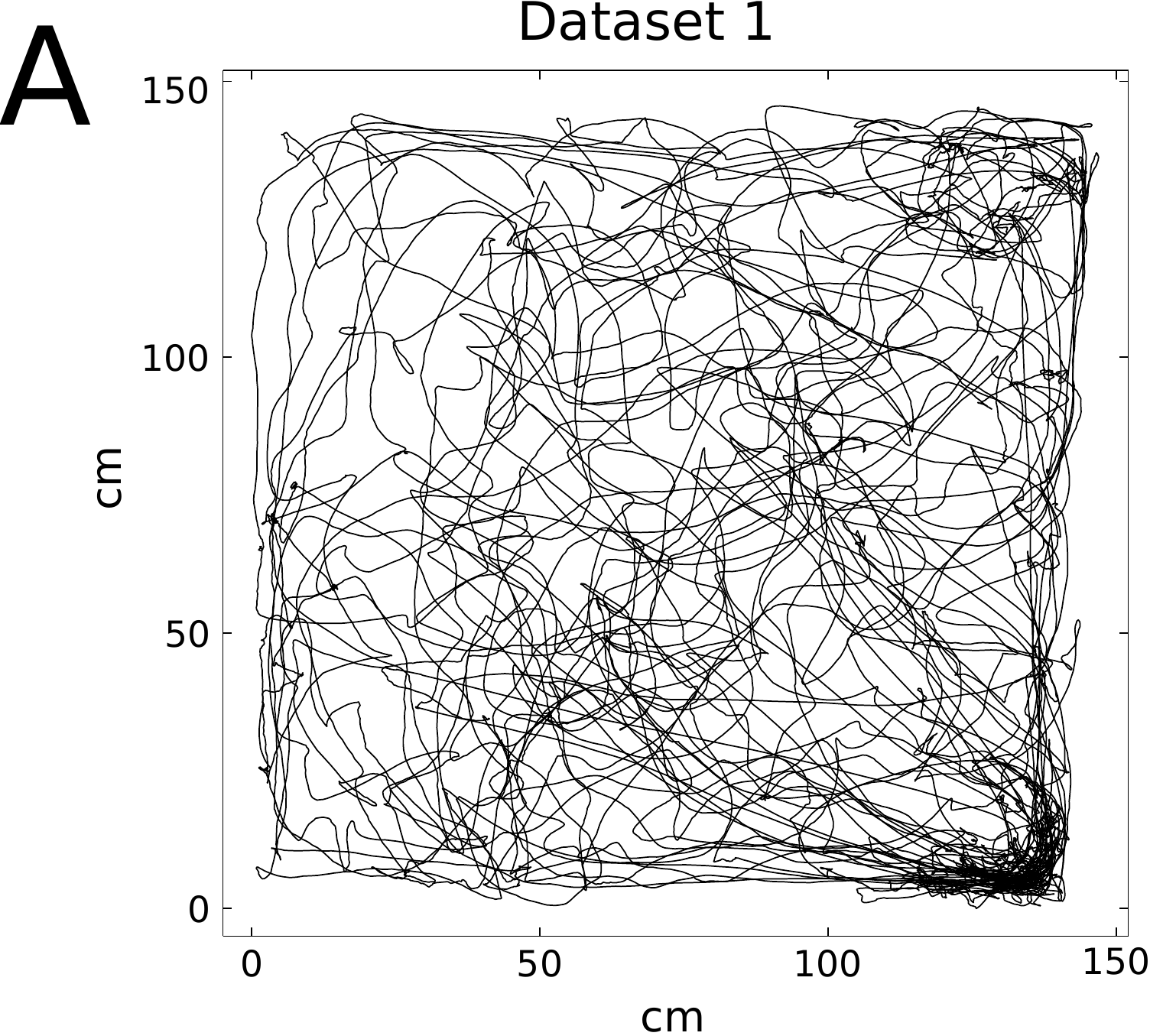} &
      \includegraphics[width=44.5mm]{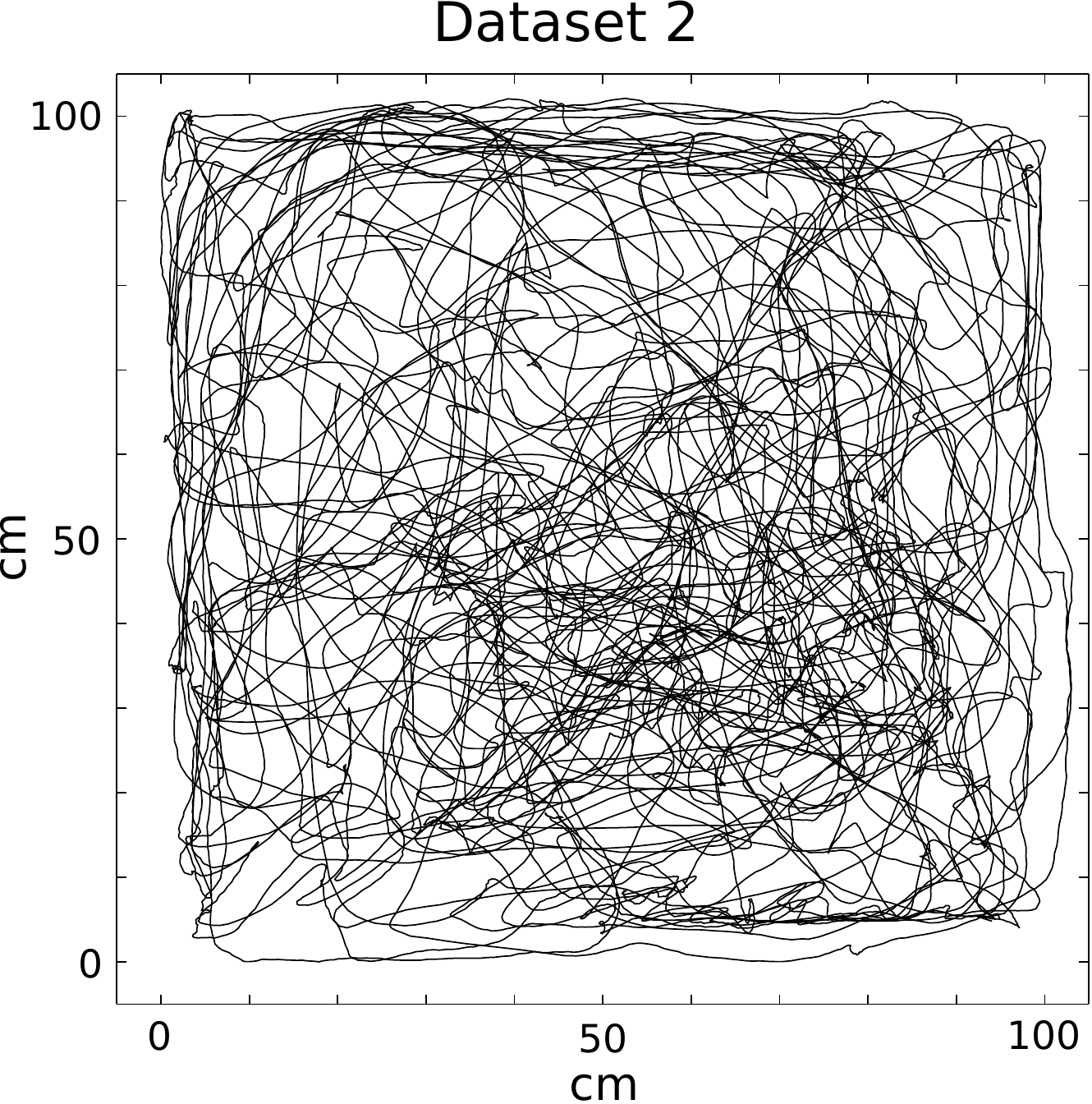} \\
      \includegraphics[width=50mm]{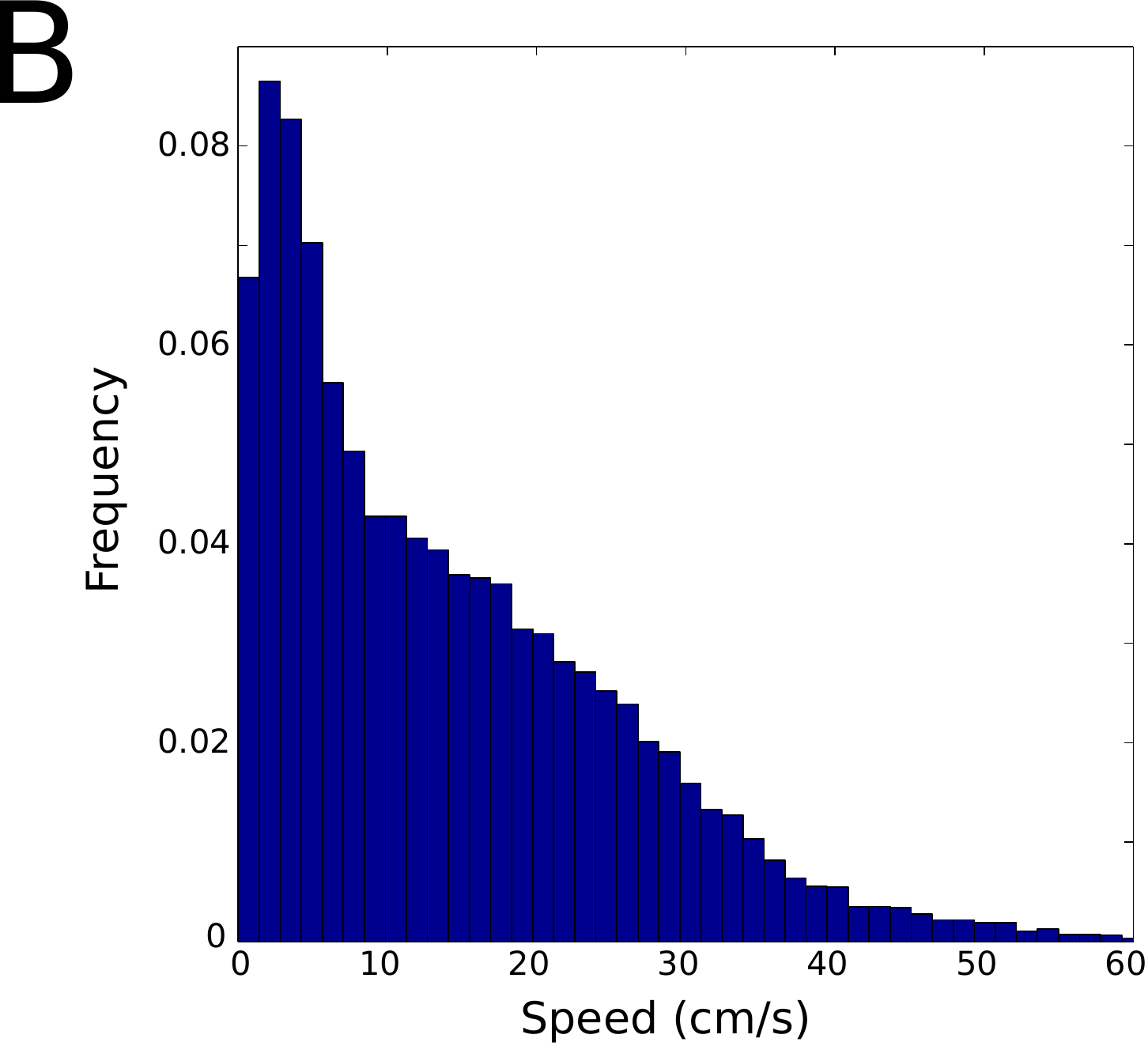} &            
      \includegraphics[width=46.5mm]{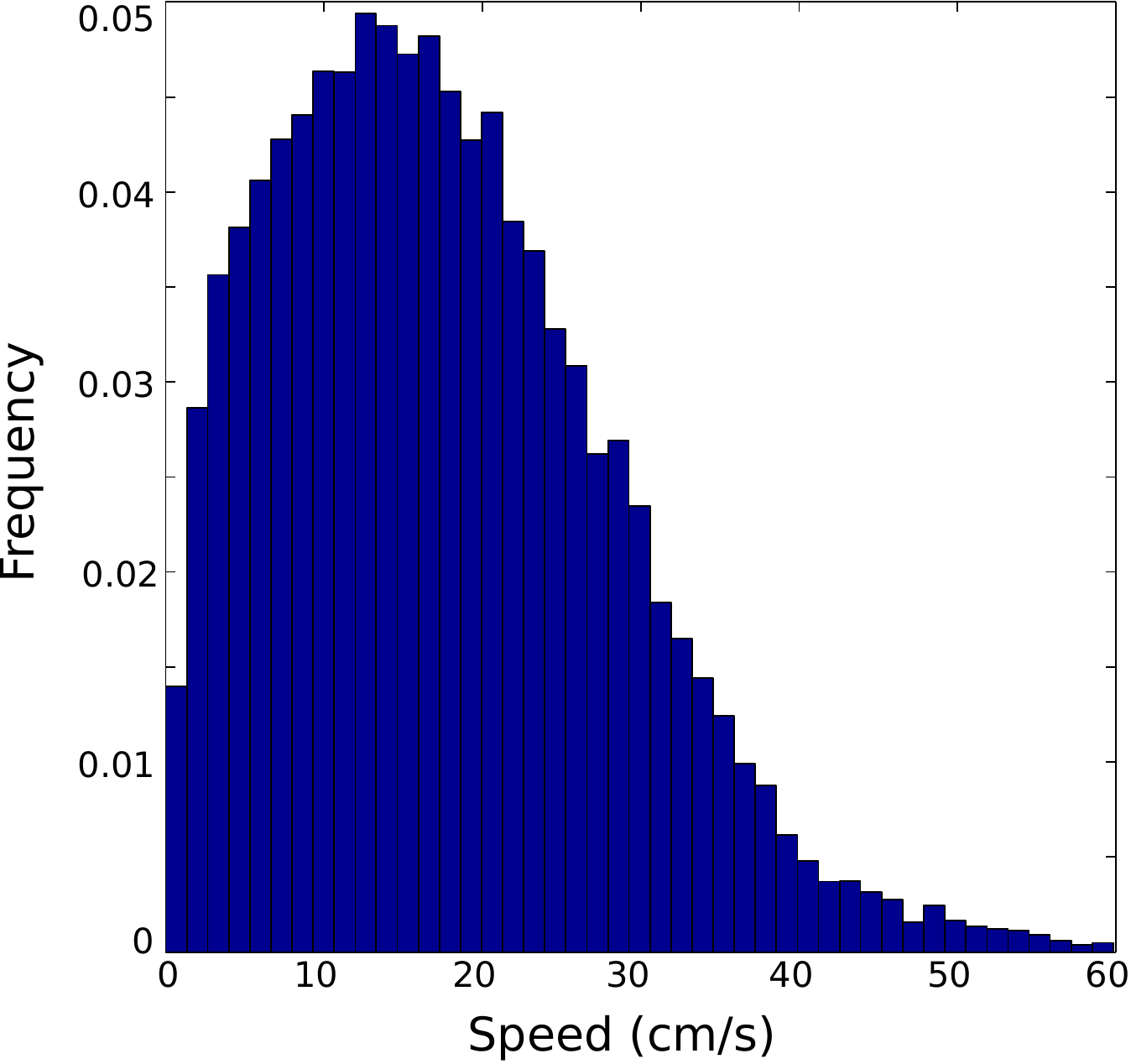}    
    \end{tabular}
  \end{center}
  \caption{\textbf{Trajectory and speed.} \small{In panel A, the trajectories of the rats are shown. 
  Figures in Panel B show the frequencies of different speeds during the recordings for the two data sets.}}
  \label{fig:Runs}
\end{figure}
 Two recordings of the activity of cells in the MEC area of two Long Evans male rats (from \cite{Stensola2012}) were analyzed in this paper. One recording, referred to as data set 1, 
consisted of a total of 65 cells (rat 14147 in \cite{Stensola2012}), where 27 were classified as grid cells (mean firing rate: 2.4 Hz). These 27 cells distributed over 7 tetrodes, and 22 of them could 
be assigned to one of three modules (see \cite{Stensola2012} for methods). The number of cells in each module, along with mean spacing and orientation is given in 
Table \ref{tab:Modules}. The other recording, data set 2, consisted of 8 grid cells (mean firing rate: 2.8 Hz) distributed over 3 tetrodes (rat 13855 in \cite{Stensola2012}). 
All 8 cells belonged to the same module. Mean spacing and orientation for this module is listed in Table \ref{tab:Modules}. 
The movement of the rats is shown in Fig.\ \ref{fig:Runs}.
\begin{center}

\captionof{table}{Mean spacing and orientation for the 3+1 modules} 
\label{tab:Modules} 
\begin{tabular}{|c|c|c|c|} \hline
& Module & Mean spacing$\pm$std & Mean orientation$\pm$std \\ \hline
Data set 1 & 1 (8 cells) & 46.4$\pm$1.7cm & 31.5$\pm$1.9\textdegree\\ \cline{2-4}
& 2 (7 cells) & 46.4$\pm$1.7cm & 27.6$\pm$2.7\textdegree \\ \cline{2-4}
& 3 (7 cells) & 93.2$\pm$2.6cm & 35.0$\pm$2.4\textdegree \\ \hline
Data set 2 & 1 (8 cells) & 31.1$\pm$1.3cm & 14.7$\pm$1.6\textdegree\\ \hline 
\end{tabular}
\end{center}

The spikes were binned into 10 ms time bins, but using both 20 ms and 5 ms time bins led to similar results.
Using the binned data, a spike matrix of -1's and 1's was constructed, where a `-1' indicated that the cell did not
fire in time bin t, and a `1' indicated that the cell emitted one or more spikes in time bin t. More than one spike rarely happened (both data sets: average over cells = 0.1 ($\pm$0.1) \% of the time bins).

\subsection*{Noise correlations}

Noise correlations were defined as
\begin{equation*}
C_{ij} = \langle \rho(\bar{\textbf{r}}_i^{a}, \bar{\textbf{r}}_j^{a}) \rangle_a 
\end{equation*}
where $\bar{\textbf{r}}_i^{a}$ is a 1$\times{}k$ vector consisting of the average firing rate of neuron $i$  in each of the $k$ 
trajectories through spatial bin $a$, and $\rho(\cdot,\cdot)$ is the Pearson correlation coefficient (PCC), defined as:

\begin{equation*}
  \rho(\bar{\textbf{r}}_i^{a},\bar{\textbf{r}}_j^{a}) = \frac{E\big[(\bar{\textbf{r}}_i^{a}-\langle \bar{\textbf{r}}_i^{a} \rangle_k)(\bar{\textbf{r}}_j^{a}-\langle \bar{\textbf{r}}_j^{a} \rangle_k   ) \big]}{\sigma[{{\textbf{r}}_i^{a}}] \times \sigma[{{\textbf{r}}_j^{a}}]}
\end{equation*}
with both the expectation ($E$) and the standard deviations ($\sigma$) over the $k$ trajectories.

Random partitions: Each spatial bin has a given number of visits. To split the data into two 
random partitions, for all visited bins, a randomly chosen half of the visits to each bin was assigned to one partition, 
the other half to the other partition.

\subsection*{Theta clustering}
\begin{figure}[h!]
  \begin{center}
     \includegraphics[width=0.38\textwidth,natwidth=776,natheight=1024]{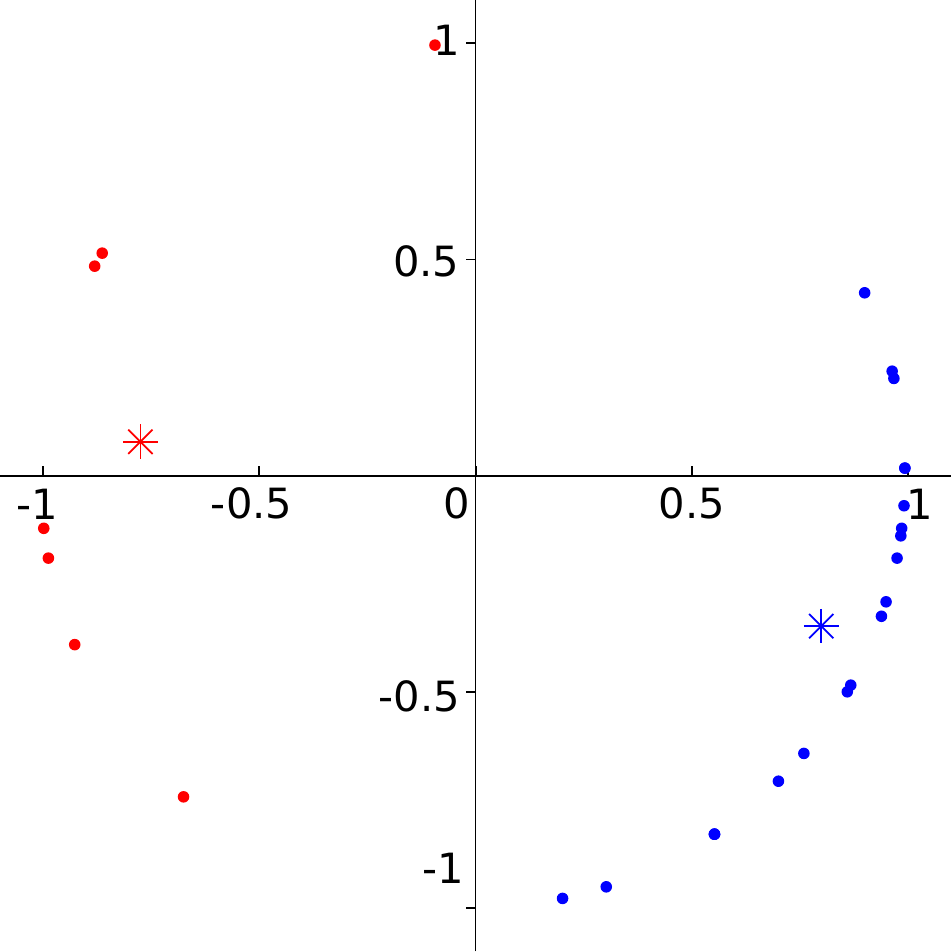}
  \end{center}
  \caption{\textbf{Theta clusters.} \small{Preferred phase of theta of the 27 cells in data set 1 plotted onto the unit circle. 
  The cells were clustered into two groups (red and blue) (see Material and Methods). Asterisks mark cluster centers.}
  \label{fig:NCandThetaClusters}}
\end{figure}
The cells could be divided in two clusters based on preferred phase of theta. The
theta phase preference was defined as the peak in a circular kernel smoothed density estimate of the
distribution of theta value at spike time. The number of clusters were defined as the number of local peaks 
in a kernel smoothed density estimate of the distribution of theta phase preference peaks for all cells.
A circular $k$-means clustering algorithm were performed to assign cells to clusters. The clusters are shown in Fig.\ \ref{fig:NCandThetaClusters}.

\subsection*{Model definition and inference}
We used the kinetic Ising model to infer the functional network connectivity, i.e.
we assumed that the observed spike train comes from the probability
distribution in Eq. \ref{eq:kineticIsingPdist}.
We constructed different versions of the model by varying the form of the external field in several ways as described in the introduction and in more details below.

To allow the external field of the kinetic Ising model to account for the
spatial variations in the firing of the grid cells, we started, for data set 1, by dividing
the environment globally into K square boxes. We defined three models
with increasing spatial resolution, with K = $4 \times 4$ (37.5 cm boxes) in the first
model, and K = $20 \times 20$ (7.5 cm boxes) in the second.
For each K, we defined external fields $\alpha_{ik}$ for each cell $i$ and box $k$.
The field resulting from this spatial discretization is then
$h_i^{\text{S}}$(t) $= \sum\nolimits_k \alpha_{ik} I_{k}(t)$, where $I_{k}(t)$
is a function indicating the presence ($1$) or absence ($0$) of the animal
in box $k$ at time $t$.

We further increased the resolution of the spatial fields using
Gaussian basis functions centered on an evenly spaced $M \times M$ square lattice covering the recording environment.
The spatial field for cell $i$ at time $t$ is then
\begin{equation}
h^{\text{S}}_i(t) = \sum\nolimits_{jk}{\alpha_{ijk}\exp\big[-\big((x(t)-x_{jk})^2 + (y(t)-y_{jk})^2 \big) / r^2}\big] + h_i
\end{equation}
where ($x_{jk}$,$y_{jk}$)  and $r$ are the vertices of the regular lattice and the widths of the basis functions, respectively.
To determine the optimal values of $M$ and $r$ ($M=15$ and $r=8.5$ cm), we maximized the likelihood for a range of values of $M$ and $r$ 
and chose the values of the parameters that gave the highest Akaike-adjusted likelihood value.
To include the external theta phase preference, we computed the fast-Fourier transform of the local field potential (LFP) and set the 
theta rhythm to the maximum component between 4-12 Hz.
From this, we constructed a theta input vector, where each element was the angular average $\in (-\pi, \pi]$ of the theta phase in that time bin.
The partial field for cell $i$ at time $t$ due to local field potential theta preference is then
\begin{equation}
h^{\text{LFP}}_i(t) = \sum\nolimits_{k} \alpha_{ik}\exp\big[- d(\Theta(t),\Theta_{k})^2 /(\pi/6)^2 \big] + h_i
\label{eq:theta}
\end{equation}
where $d(\Theta(t), \Theta_k$) is the minimum angular distance between $\Theta(t)$, the theta phase in time
bin $t$, and $\Theta_k$, the k'th component of a set of 10 equally spaced angular phases. 
The number of angles and width of Gaussian ($\pi/6$) was selected by maximizing the Akaike-adjusted
likelihood of the model in the same way parameter values for $M$ and $r$ in the model with spatial fields were chosen, as described above.

The head and running direction components was also accounted for using sums of Gaussian basis functions
\begin{equation}
h^{\text{HD}}_i(t) = \sum\nolimits_{k}{\alpha_{ik}\exp\big[-{d(\phi(t), \phi_{k})^2} /{({\pi}/{6})^2}}\big] + h_i
\label{eq:HD}
\end{equation}
where $\phi(t)$ is the head direction $\in (-\pi, \pi]$ at time $t$, calculated from the projection of 
two LEDs onto the horizontal plane, and $\phi_k$ is the angular position of
the kth basis function. The number of basis functions (10) and width of Gaussian ($\pi/6$) were selected by maximizing the Akaike-adjusted
likelihood of the model, the same way it was done for parameter choice in the spatial and theta model. 
Speed was also incorporated into the model with a simple time-varying field, $\alpha_i s(t)$, where
 $s(t)$ is the average speed in the 100ms window around each time bin.

In all of the models, the parameters, $J_{ij}$, $h_i$ and $\alpha$'s, were found by
maximizing the likelihood function given in (\ref{eq:likelihood}) for the data under the different models
by gradient ascent.  When comparing the models, we first Akaike-corrected the log-likelihood. The Akaike information criterion (AIC) is a 
measure to compensate for overfitting by models with more parameters, where the preferred model is that with the minimum AIC value,  
defined as
\begin{equation}
AIC = -2 \ln\big(L[\mathbf{D}|\boldsymbol{\theta}_{ML}]\big) + 2k
\label{eq:AIC}
\end{equation}
where \textbf{D} is the observed data, and $L[\mathbf{D}|\boldsymbol{\theta}_{ML}]$ is the likelihood at the 
maximum likelihood (ML) estimates of the parameters $\boldsymbol{\theta}$ ($\boldsymbol{\theta}_{ML}$), and $k$ is the number of parameters \cite{Akaike}. 
Equivalent to the method described above, we corrected the total log-likelihood as ln$\big(L_{\text{Akaike}}\big) = -\frac{AIC}{2}$.

\section*{Acknowledgements}
We are most grateful to Hanne and Tor Stensola for providing the data. We thank John Hertz, Dori Derdikman, Edvard and May-Britt Moser for valueable feedback on this work.

\bibliography{myrefs}

\begin{thebibliography}{10}
\providecommand{\url}[1]{\texttt{#1}}
\providecommand{\urlprefix}{URL }
\expandafter\ifx\csname urlstyle\endcsname\relax
  \providecommand{\doi}[1]{doi:\discretionary{}{}{}#1}\else
  \providecommand{\doi}{doi:\discretionary{}{}{}\begingroup
  \urlstyle{rm}\Url}\fi
\providecommand{\bibAnnoteFile}[1]{%
  \IfFileExists{#1}{\begin{quotation}\noindent\textsc{Key:} #1\\
  \textsc{Annotation:}\ \input{#1}\end{quotation}}{}}
\providecommand{\bibAnnote}[2]{%
  \begin{quotation}\noindent\textsc{Key:} #1\\
  \textsc{Annotation:}\ #2\end{quotation}}
\providecommand{\eprint}[2][]{\url{#2}}

\bibitem{Hafting2005}
Hafting T, Fyhn M, Molden S, Moser M, Moser E (2005) Microstructure of a
  spatial map in the entorhinal cortex.
\newblock Nature 436: 801-6.
\bibAnnoteFile{Hafting2005}

\bibitem{Fyhn2004}
Fyhn M, Molden S, Witter MP, Moser EI, Moser MB (2004) Spatial representation
  in the entorhinal cortex.
\newblock Science 305: 1258-64.
\bibAnnoteFile{Fyhn2004}

\bibitem{Sargolini2006}
Sargolini F, Fyhn M, Hafting T, McNaughton BL, Witter MP, et~al. (2006)
  Conjunctive representation of position, direction, and velocity in entorhinal
  cortex.
\newblock Science 312: 758-62.
\bibAnnoteFile{Sargolini2006}

\bibitem{Yartsev2011}
Yartsev MM, Witter MP, Ulanovsky N (2011) Grid cells without theta oscillations
  in the entorhinal cortex of bats.
\newblock Nature 479: 103-107.
\bibAnnoteFile{Yartsev2011}

\bibitem{killian2012map}
Killian NJ, Jutras MJ, Buffalo EA (2012) A map of visual space in the primate
  entorhinal cortex.
\newblock Nature 491: 761--764.
\bibAnnoteFile{killian2012map}

\bibitem{jacobs2013direct}
Jacobs J, Weidemann CT, Miller JF, Solway A, Burke JF, et~al. (2013) Direct
  recordings of grid-like neuronal activity in human spatial navigation.
\newblock Nat neurosci 16: 1188--1190.
\bibAnnoteFile{jacobs2013direct}

\bibitem{Stensola2012}
Stensola H, Stensola T, Solstad T, Fr\o{}land K, Moser MB, et~al. (2012) The
  entorhinal grid map is discretized.
\newblock Nature 492: 72-8.
\bibAnnoteFile{Stensola2012}

\bibitem{moser2014network}
Moser EI, Moser MB, Roudi Y (2014) Network mechanisms of grid cells.
\newblock Philos Trans R Soc Lond B Biol Sci 369: 20120511.
\bibAnnoteFile{moser2014network}

\bibitem{moser2014grid}
Moser EI, Roudi Y, Witter MP, Kentros C, Bonhoeffer T, et~al. (2014) Grid cells
  and cortical representation.
\newblock Nature Reviews Neuroscience .
\bibAnnoteFile{moser2014grid}

\bibitem{yoon2013specific}
Yoon K, Buice MA, Barry C, Hayman R, Burgess N, et~al. (2013) Specific evidence
  of low-dimensional continuous attractor dynamics in grid cells.
\newblock Nat neurosci 16: 1077--1084.
\bibAnnoteFile{yoon2013specific}

\bibitem{McNaughton2006}
McNaughton BL, Battaglia FP, Jensen O, Moser EI, Moser MB (2006) Path
  integration and the neural basis of the'cognitive map'.
\newblock Nat Rev Neurosci 7: 663--678.
\bibAnnoteFile{McNaughton2006}

\bibitem{Fuhs2006}
Fuhs MC, Touretzky DS (2006) A spin glass model of path integration in rat
  medial entorhinal cortex.
\newblock J Neurosci 26: 4266--4276.
\bibAnnoteFile{Fuhs2006}

\bibitem{Burak2009}
Burak Y, Fiete IR (2009) Accurate path integration in continuous attractor
  network models of grid cells.
\newblock PLoS Comput Biol 5: e1000291.
\bibAnnoteFile{Burak2009}

\bibitem{Couey2013}
Couey JJ, Witoelar A, Zhang SJ, Zheng K, Ye J, et~al. (2013) Recurrent
  inhibitory circuitry as a mechanism for grid formation.
\newblock Nat Neurosci 16: 318-24.
\bibAnnoteFile{Couey2013}

\bibitem{Truccolo2005}
Truccolo W, Eden UT, Fellows MR, Donoghue JP, Brown EN (2005) A point process
  framework for relating neural spiking activity to spiking history, neural
  ensemble, and extrinsic covariate effects.
\newblock J Neurophysiol 93: 1074--1089.
\bibAnnoteFile{Truccolo2005}

\bibitem{pillow2008spatio}
Pillow JW, Shlens J, Paninski L, Sher A, Litke AM, et~al. (2008)
  Spatio-temporal correlations and visual signalling in a complete neuronal
  population.
\newblock Nature 454: 995--999.
\bibAnnoteFile{pillow2008spatio}

\bibitem{Rebesco2010}
Rebesco JM, Stevenson IH, K{\"o}rding KP, Solla SA, Miller LE (2010) Rewiring
  neural interactions by micro-stimulation.
\newblock Front Syst Neurosci 4.
\bibAnnoteFile{Rebesco2010}

\bibitem{mathis2013multiscale}
Mathis A, Herz AVM, Stemmler MB (2013) Multiscale codes in the nervous system:
  The problem of noise correlations and the ambiguity of periodic scales.
\newblock Phys Rev E 88.
\bibAnnoteFile{mathis2013multiscale}

\bibitem{nelder1972generalized}
Nelder JA, Baker R (1972) Generalized linear models.
\newblock Wiley Online Library.
\bibAnnoteFile{nelder1972generalized}

\bibitem{jaynes1982rationale}
Jaynes ET (1982) On the rationale of maximum-entropy methods.
\newblock Proceedings of the IEEE 70: 939--952.
\bibAnnoteFile{jaynes1982rationale}

\bibitem{schneidman2006weak}
Schneidman E, Berry MJ, Segev R, Bialek W (2006) Weak pairwise correlations
  imply strongly correlated network states in a neural population.
\newblock Nature 440: 1007--1012.
\bibAnnoteFile{schneidman2006weak}

\bibitem{shlens2006structure}
Shlens J, Field GD, Gauthier JL, Grivich MI, Petrusca D, et~al. (2006) The
  structure of multi-neuron firing patterns in primate retina.
\newblock J Neurosci 26: 8254--8266.
\bibAnnoteFile{shlens2006structure}

\bibitem{Roudi2011meanfield}
Roudi Y, Hertz J (2011) Mean field theory for nonequilibrium network
  reconstruction.
\newblock Phys Rev Lett 106: 048702.
\bibAnnoteFile{Roudi2011meanfield}

\bibitem{jeewajee2014theta}
Jeewajee A, Barry C, Douchamps V, Manson D, Lever C, et~al. (2014) Theta phase
  precession of grid and place cell firing in open environments.
\newblock Philosophical Transactions of the Royal Society B: Biological
  Sciences 369: 20120532.
\bibAnnoteFile{jeewajee2014theta}

\bibitem{ecker2010decorrelated}
Ecker AS, Berens P, Keliris GA, Bethge M, Logothetis NK, et~al. (2010)
  Decorrelated neuronal firing in cortical microcircuits.
\newblock Science 327: 584--587.
\bibAnnoteFile{ecker2010decorrelated}

\bibitem{ventura2012accurately}
Ventura V, Gerkin RC (2012) Accurately estimating neuronal correlation requires
  a new spike-sorting paradigm.
\newblock Proceedings of the National Academy of Sciences 109: 7230--7235.
\bibAnnoteFile{ventura2012accurately}

\bibitem{gray1995tetrodes}
Gray CM, Maldonado PE, Wilson M, McNaughton B (1995) Tetrodes markedly improve
  the reliability and yield of multiple single-unit isolation from multi-unit
  recordings in cat striate cortex.
\newblock Journal of neuroscience methods 63: 43--54.
\bibAnnoteFile{gray1995tetrodes}

\bibitem{Akaike}
Akaike H (1974) A new look at the statistical model identification.
\newblock IEEE T Automat Contr 19: 716--723.
\bibAnnoteFile{Akaike}

\bibitem{pastoll2013feedback}
Pastoll H, Solanka L, van Rossum MC, Nolan MF (2013) Feedback inhibition
  enables theta-nested gamma oscillations and grid firing fields.
\newblock Neuron 77: 141--154.
\bibAnnoteFile{pastoll2013feedback}

\bibitem{beed2013inhibitory}
Beed P, Gundlfinger A, Schneiderbauer S, Song J, B{\"o}hm C, et~al. (2013)
  Inhibitory gradient along the dorsoventral axis in the medial entorhinal
  cortex.
\newblock Neuron 79: 1197--1207.
\bibAnnoteFile{beed2013inhibitory}

\bibitem{Dhillon2000}
Dhillon A, Jones RS (2000) Laminar differences in recurrent excitatory
  transmission in the rat entorhinal cortex $in$ $vitro$.
\newblock Neuroscience 99: 413--422.
\bibAnnoteFile{Dhillon2000}

\bibitem{buetfering2014parvalbumin}
Buetfering C, Allen K, Monyer H (2014) Parvalbumin interneurons provide grid
  cell-driven recurrent inhibition in the medial entorhinal cortex.
\newblock Nature neuroscience .
\bibAnnoteFile{buetfering2014parvalbumin}

\bibitem{lezon2006using}
Lezon TR, Banavar JR, Cieplak M, Maritan A, Fedoroff NV (2006) Using the
  principle of entropy maximization to infer genetic interaction networks from
  gene expression patterns.
\newblock Proc Natl Acad Sci U S A 103: 19033--19038.
\bibAnnoteFile{lezon2006using}

\bibitem{Fyhn2007}
Fyhn M, Hafting T, Treves A, Moser MB, Moser EI (2007) Hippocampal remapping
  and grid realignment in entorhinal cortex.
\newblock Nature 446: 190--194.
\bibAnnoteFile{Fyhn2007}

\bibitem{burgess2007oscillatory}
Burgess N, Barry C, O'Keefe J (2007) An oscillatory interference model of grid
  cell firing.
\newblock Hippocampus 17: 801--812.
\bibAnnoteFile{burgess2007oscillatory}

\bibitem{kropff2008emergence}
Kropff E, Treves A (2008) The emergence of grid cells: Intelligent design or
  just adaptation?
\newblock Hippocampus 18: 1256--1269.
\bibAnnoteFile{kropff2008emergence}

\bibitem{si2012grid}
Si B, Kropff E, Treves A (2012) Grid alignment in entorhinal cortex.
\newblock Biological cybernetics 106: 483--506.
\bibAnnoteFile{si2012grid}

\bibitem{Hertz2011}
Hertz J, Roudi Y, Tyrcha J (2011) Ising models for inferring network structure
  from spike data.
\newblock arXiv:11061752 [q-bioQM] .
\bibAnnoteFile{Hertz2011}

\bibitem{capone2014inferring}
Capone C, Filosa C, Gigante G, Ricci-Tersenghi F, del Giudice P (2014)
  Inferring synaptic structure in presence of neural interaction time scales.
\newblock arXiv preprint arXiv:14081015 .
\bibAnnoteFile{capone2014inferring}

\bibitem{Roudi2009}
Roudi Y, Nirenberg S, Latham PE (2009) Pairwise maximum entropy models for
  studying large biological systems: when they can work and when they can't.
\newblock PLoS Comput Biol 5: e1000380.
\bibAnnoteFile{Roudi2009}

\bibitem{stevenson2012functional}
Stevenson IH, London BM, Oby ER, Sachs NA, Reimer J, et~al. (2012) Functional
  connectivity and tuning curves in populations of simultaneously recorded
  neurons.
\newblock PLoS Comput Biol 8: e1002775.
\bibAnnoteFile{stevenson2012functional}

\end{thebibliography}
\end{document}